\documentclass[
 reprint,
superscriptaddress,
 amsmath,amssymb,
 aps,
pra,
]{revtex4-1}

\usepackage{graphicx}
\usepackage{dcolumn}
\usepackage{bm}
\usepackage{hyperref}

\begin{document}

\title{Decoding Quantum Error Correction Codes with Local Variation}

\author{Michael Hanks}
\affiliation{Department of Informatics, School of Multidisciplinary Sciences, Sokendai (The Graduate University for Advanced Studies), 2-1-2 Hitotsubashi, Chiyoda-ku, Tokyo 101-8430 Japan}
\affiliation{National Institute of Informatics, 2-1-2 Hitotsubashi, Chiyoda-ku, Tokyo 101-8430, Japan}

\author{William J. Munro}
\affiliation{NTT Basic Research Laboratories \& NTT Research Center for Theoretical Quantum Physics, NTT Corporation, 3-1 Morinosato-Wakamiya, Atsugi, Kanagawa 243-0198, Japan}
\affiliation{National Institute of Informatics, 2-1-2 Hitotsubashi, Chiyoda-ku, Tokyo 101-8430, Japan}
\author{Kae Nemoto}
\affiliation{National Institute of Informatics, 2-1-2 Hitotsubashi, Chiyoda-ku, Tokyo 101-8430, Japan}
\affiliation{Department of Informatics, School of Multidisciplinary Sciences, Sokendai (The Graduate University for Advanced Studies), 2-1-2 Hitotsubashi, Chiyoda-ku, Tokyo 101-8430 Japan}

\date{\today}

\begin{abstract}
In this paper we investigate the role of local information in the decoding of the \emph{repetition} and \emph{surface} error correction codes for the protection of quantum states. Our key result is an improvement in resource efficiency when local information is taken into account during the decoding process: the \emph{code distance} associated with a given logical error rate is reduced with a magnitude depending on the proximity of the physical error rate to the \emph{accuracy threshold} of the code. We also briefly discuss an averaged approach with local information for table-lookup and localised decoding schemes, an expected breakdown of these effects for large-scale systems, and the importance of this resource reduction in the near-term.
\end{abstract}

\maketitle

\section{Introduction}
\label{sec:introduction}

It has long been known that quantum information processing devices at any significant scale will face the obstacles of cumulative noise and error \cite{Landauer95a,Chuang95a, Unruh95a, Palma96a}. Quantum error correction codes \cite{Lidar13a,Devitt13a,Terhal15a} were developed to overcome these obstacles, at the cost of increased resource (qubit) and time overheads. Even for smaller devices in the near-term, partially error-corrected approaches have been proposed to mitigate limiting noise processes \cite{Arrad14a, Kessler14a, Duer14a, Unden16a}. Present devices face tight resource restrictions and error rates comparable to even the largest accuracy thresholds, so it is not (yet) sufficient to treat quantum error correction schemes as if their choice was agnostic with respect to the underlying technology or application: we must consider all idiosyncrasies and constraints before us. The constraints of most physical systems mean that the family of topological quantum error correction codes seems most promising for any near- to mid-term development, having three characteristic advantages: large accuracy thresholds, small correction circuits, and local interactions. The surface code \cite{Bravyi98a,Freedman01a,Dennis02a}, for example, requires only nearest-neighbour interactions. Conversely, many codes without such local constraints, such as Shor's code \cite{Shor95a} and other concatenated codes \cite{Knill96a}, are simply out of reach for many real physical systems.

Just as the form of interaction varies according to our choice of physical system, so too does the form of the noise and error we confront; the very earliest proposals for quantum error correction in fact relied on error \emph{detection} schemes \cite{Berthiaume94a,Bennett94a,Vaidman96a,Barenco97a}, under the assumption that the error of the state was subject to the quantum Zeno effect \cite{Misra77a}. Standard models did eventually settle on the \emph{depolarising noise channel} \cite{Bennett96a}, but even then parallel streams of development emerged to deal with quantum channels for which depolarising noise was insufficient, such as loss channels \cite{Ralph05a}. In the last decade we have seen a plethora of results looking at different noise models, verifying the performance of the codes under such models and asking what modifications, if any, might be made to improve performance. Early examples focussed on the tendency for errors to be highly \emph{biased} toward one particular basis (such as dephasing) \cite{Aliferis08a,Brooks13a,Stephens13a}. Investigations of qubit loss \cite{Stace09a,Stace10a,Barrett10a,Ohzeki12a,Fujii12a}, amplitude damping \cite{Tomita14a,Darmawan17a}, correlation \cite{Wang11a,Bombin12a,Jouzdani13a,Novais13a,Jouzdani14a,Jouzdani14b}, and qubit leakage \cite{Fowler13a,Suchara15a} have since been undertaken.

In this work we focus not on qualitatively distinct channel behaviour, such as loss or amplitude damping, but on local variation in a standard depolarising noise channel. Specifically, we assume that the measurement outcomes associated with each stabiliser operation may be distinct with respect to their information content. This variability will not be the result of any changing external influence, but inherent in the information content associated with the measurement outcomes themselves. {Any multi-shot \cite{Hanks17a} or long-time count-threshold \cite{Bhaskar19a} measurement scheme in the presence of error is expected to display such local variation.} We perform pseudo-threshold simulations for the repetition and surface codes, comparing the standard, fixed-error-rate {phenomenological error model with the case for} an error-rate drawn from a discrete, balanced, two-component distribution $D$ of equal mean $p_{\mu}$ but a fixed relative width $\sigma$,
\begin{align}
	D(x;p_{\mu},\sigma) &= \frac{\delta\left[ x-p_{\mu}(1-\sigma)\right] + \delta\left[x-p_{\mu}(1+\sigma)\right]}{2},\label{eq:sigmaBinaryDistributionDefinition}
\end{align}
where $\delta[\cdot]$ is the delta function {and $p_{\mu}$ will serve as the phenomenological physical error rate \cite{Wang03a,Stephens14a} in addition to the mean measurement error rate}. This toy distribution is chosen to maximise the contrast between different sites, to accentuate the effects of the variability, and because the kind of feedback in measurement that we envisage is expected to result in a discrete distribution (as opposed to continuous alternatives such as the Normal Distribution). We also introduce two approximate measures consistent with the simulated results to extrapolate the significance of variation for larger codes and higher dimensions. Our paper is organised as follows: In Section~\ref{sec:quantifying_significance} we describe and justify the approximate measures we introduce. Section~\ref{sub:the_repetition_code} then defines and takes the repetition code as an exemplar of the significance of local variation for increasing code distance, while Section~\ref{sub:the_surface_code} extends the analysis to the surface code for comparative inference about the behaviour of codes in higher dimensions. In Section~\ref{sub:discussion} we summarise our results and discuss a potential generalisation for alternative decoding schemes before concluding.

\section{Quantifying Significance}
\label{sec:quantifying_significance}

It is important to investigate the impact of local measurement variablility ($\sigma$, for our bimodal model) on the error rate as a function of the code size and structure. This will be addressed in two ways: Firstly, numerical pseudo-threshold simulations will be performed for the repetition and surface codes, allowing us to compare the \emph{logical error rate} between these two codes and across a range of code distances and local error rates. Secondly, we explain the observed numerical behaviour by modelling the probability that local variance allows an error chain of linear dimension $\lfloor{\frac{L}{2}}\rfloor$ to be less likely than one of linear dimension $\lfloor{\frac{L}{2}}\rfloor+1$, with $L$ the code distance.

\begin{figure}
  \centering
	\includegraphics[scale=0.3]{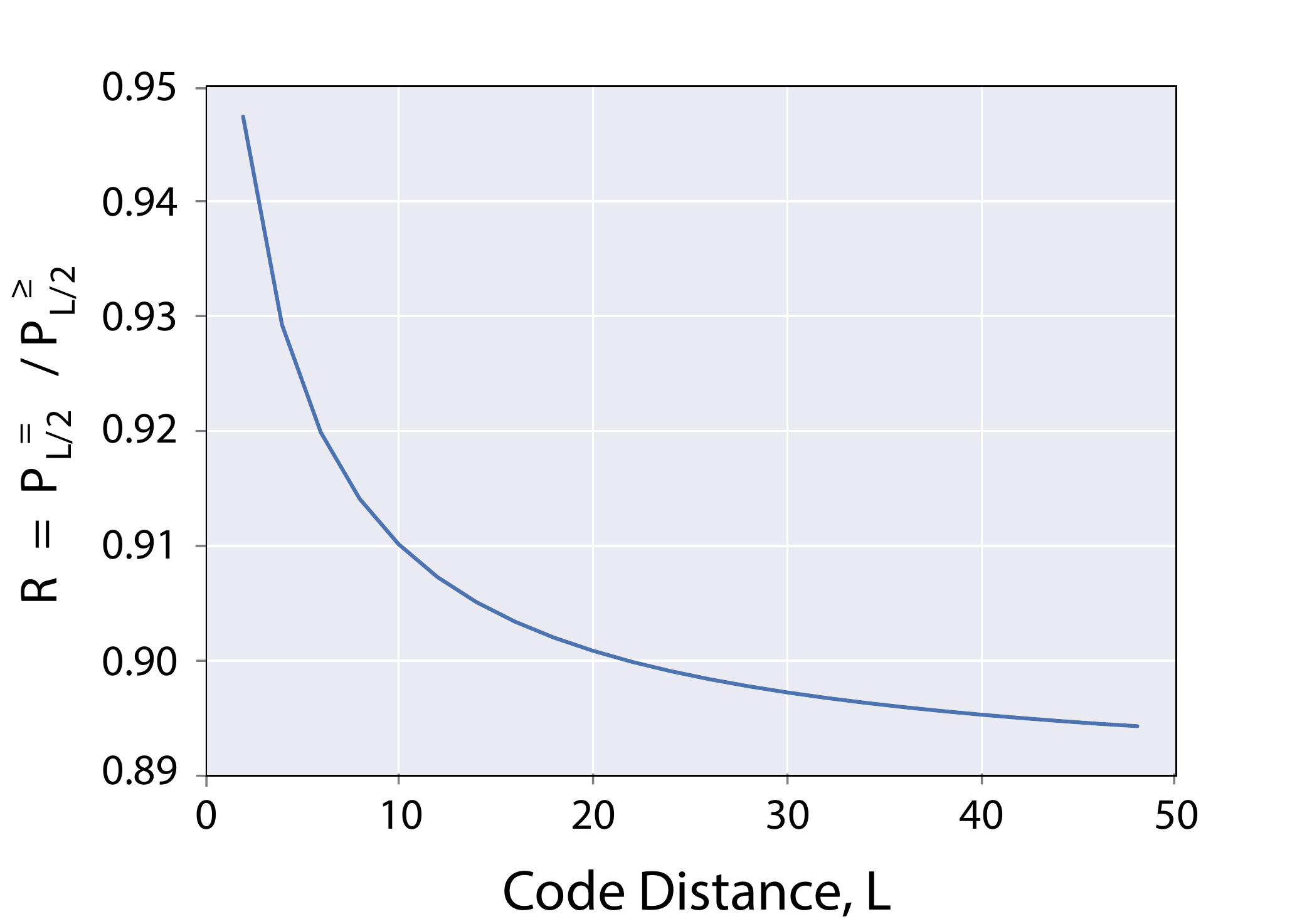}
	\caption[Fraction of Logical Errors of Weight $\lfloor{\frac{L}{2}}\rfloor$]{The fraction of logical errors caused by error chains of length $L/2$ for a one-dimensional lattice, $R=P^{\:=}_{L/2} / P^{\:\geq}_{L/2}$, as a function of the code distance $L$. The individual error rate is fixed arbitrarily at $p_{\mu}=0.1$. {The large values of this ratio ($R\geq 0.89$) indicate that the logical error rate depends significantly on our ability to distinguish between errors of adjacent lengths $L/2$ and $L/2-1$.}}
  \label{fig:noiseIntro_lOverTwoRatio}
\end{figure}

Let us first consider chains of adjacent lengths to investigate the transition point at which local information becomes useful; modifications to chains resulting in the same logical state are of at least second order in the link probability. We observe that the exponential suppression of a chain's probability with length means that local information will have the most impact when two options are close to one another in length. A chain's length is modelled as a number of successful Bernoulli trials, since a chain need not be contiguous along a given dimension of the lattice to cause a logical error. We expect chains of length $\lfloor{\frac{L}{2}}\rfloor$ will have a lesser share of the tail of this distribution if they are far above the mean number of errors $Q\cdot p_{\mu}$, where $Q$ is the total number of qubits and $p_{\mu}$ is the mean error rate per qubit (taken to be equal to the corresponding term for measurement error in Equation~\eqref{eq:sigmaBinaryDistributionDefinition}). As the code distance increases a greater fraction of logical errors should be caused by chains deviating from the half-code-distance, so that our focus on adjacent chain lengths is more valid for small code distances. {Take $P^{\:=}_{L/2}$ to be the probability of sampling an error chain of length $L/2$ from $L$ qubits via the binomial distribution,
\begin{align}
	P^{\:=}_{L/2} &=\frac{L!}{(L/2)!^{2}} \left[ p_{\mu}(1-p_{\mu}) \right]^{L/2},
\end{align}
and $P^{\:\geq}_{L/2}$ to be the probability of sampling an error chain greater than or equal to $L/2$,
\begin{align}
	P^{\:\geq}_{L/2} &= 1 - \sum^{L}_{i=\frac{L}{2}} \left(
  \begin{matrix}
    L\\
    i
  \end{matrix}
  \right) p^{i}_{\mu} (1-p_{\mu})^{L-i}.
\end{align}
Along a single dimension, we can then justify this assertion by explicitly computing the ratio of these two probabilities $R=P^{\:=}_{L/2} / P^{\:\geq}_{L/2}$,}
\begin{align}
	R &=\frac{\frac{L!}{(L/2)!^{2}} \left[ p_{\mu}(1-p_{\mu}) \right]^{L/2}}
  {1 - \sum^{L}_{i=\frac{L}{2}} \left(
  \begin{matrix}
    L\\
    i
  \end{matrix}
  \right) p^{i}_{\mu} (1-p_{\mu})^{L-i} }.\label{eq:QEC_fraction_error}
\end{align}
The behaviour of this ratio with increasing $L$ and fixed $p_{\mu}=0.1$ is shown in Figure~\ref{fig:noiseIntro_lOverTwoRatio}. Over the range of code distances considered in this paper, approximations {of relative likelihood} based on adjacent-length error chains should correspond well to true behaviour; this will be significant in { Subsections~\ref{ssub:the_impact_of_local_variance}~and~\ref{ssub:lattice_dimension},} where we develop an intuition and attempt an explanation of our observed numerical results.

\section{The Repetition Code and Chain Length}
\label{sub:the_repetition_code}

\begin{figure}[bht]
	\includegraphics[width=\linewidth]{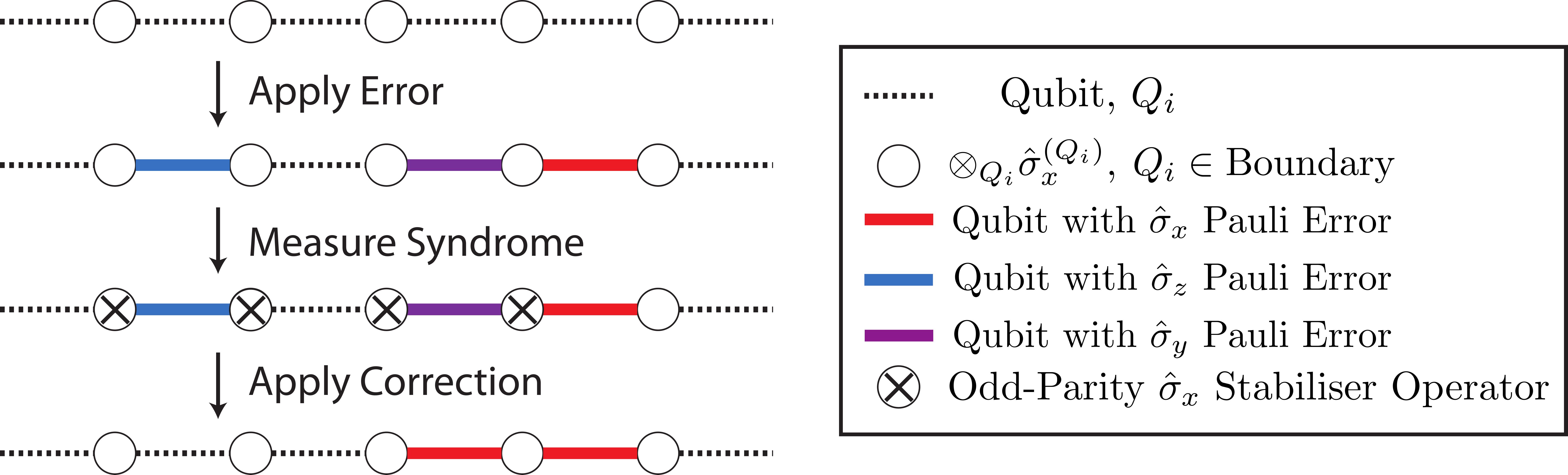}
	\caption[The Repetition Code]{A graphical representation of the repetition code. Edges represent qubits, while nodes represent $\hat{\sigma}^{(i)}_{x}\hat{\sigma}^{(i+1)}_{x}$ parity (stabiliser) operations between adjacent qubits. Errors in a basis orthogonal to the parity operations are detectable.}
  \label{fig:noiseIntro_repetition_code_diagram}
\end{figure}

The repetition code, depicted in Figure~\ref{fig:noiseIntro_repetition_code_diagram}, is defined by mapping qubit subsystems and operations to a $1\times L$ chain. It is essentially a classical code, but may nonetheless be used to partially protect quantum information and is useful when the limiting source of error is highly biased along a single dimension. The repetition code embeds one bit within the $+1$ eigenspace of parity operators $\hat{S}_{X}(v)$ acting on adjacent bits in this 1-dimensional chain,
\begin{equation}
	\label{eqn:rep_stabilizer_measurement_definition}
	\hat{S}_{X}(v) = \prod_{e | v\in\partial e} \hat{\sigma}^{(e)}_{x},
\end{equation}
where $v$ are vertices, $e$ are edges, $\partial$ denotes the boundary, and $\hat{\sigma}^{(e)}_{x}$ is the application of the Pauli X matrix,
\begin{equation}
	\label{eqn:pauli_x_matrix}
	\sigma_{x} =
	\begin{bmatrix}
		0 & 1\\
		1 & 0
	\end{bmatrix},
\end{equation}
to the qubit represented by edge $e$. Vertices of degree-$1$ are excluded. Equation~\eqref{eqn:rep_stabilizer_measurement_definition} uses a common shorthand notation for operators that are sparse with respect to the set of qubit subsystems, ignoring the order of, and trivial elements in, the tensor product in favour of superscripts.

A single local operation on any bit in the basis protected by the code (the basis orthogonal to the parity check operators) will be detected by measurement of the parity operators and may be corrected so long as the number of such errors is less than half the length of the chain. Measurement errors are incorporated by repeating parity measurements, with the effect of extending the lattice of the code into a second dimension \cite{Shor96a,Dennis02a}. The probability that accumulated error after the total set of such measurement rounds cannot be corrected is called the \emph{logical error rate}. We restrict our attention to the phenomenological error model for the duration of this report; in this model individual qubit and measurement error rates are defined \emph{per measurement round}, rather than per gate, and are associated with lattice edges.

\subsection{Time--Constant Error Rates}
\label{sub:time_constant_error_rates}

We begin by considering the simplest case where the measurement error varies spatially between sites, but is constant at each site in time; this form of error we call `static'. This is in contrast to the runtime-error to be considered in the following sections, but given the near-universality of static inhomogeneity in quantum devices, and given the computational costs of real-time decoding as systems increase in size, some distinct emphasis on this form of variation is thought useful. Inhomogeneity in detector efficiency is very common. Ranges for system detection efficiencies appear to be of order $10$\%, so even once we get the mean values down toward our target threshold, we expect significant remaining spreads.

\begin{figure}[ht]
	\begin{center}
		\includegraphics[scale=0.3]{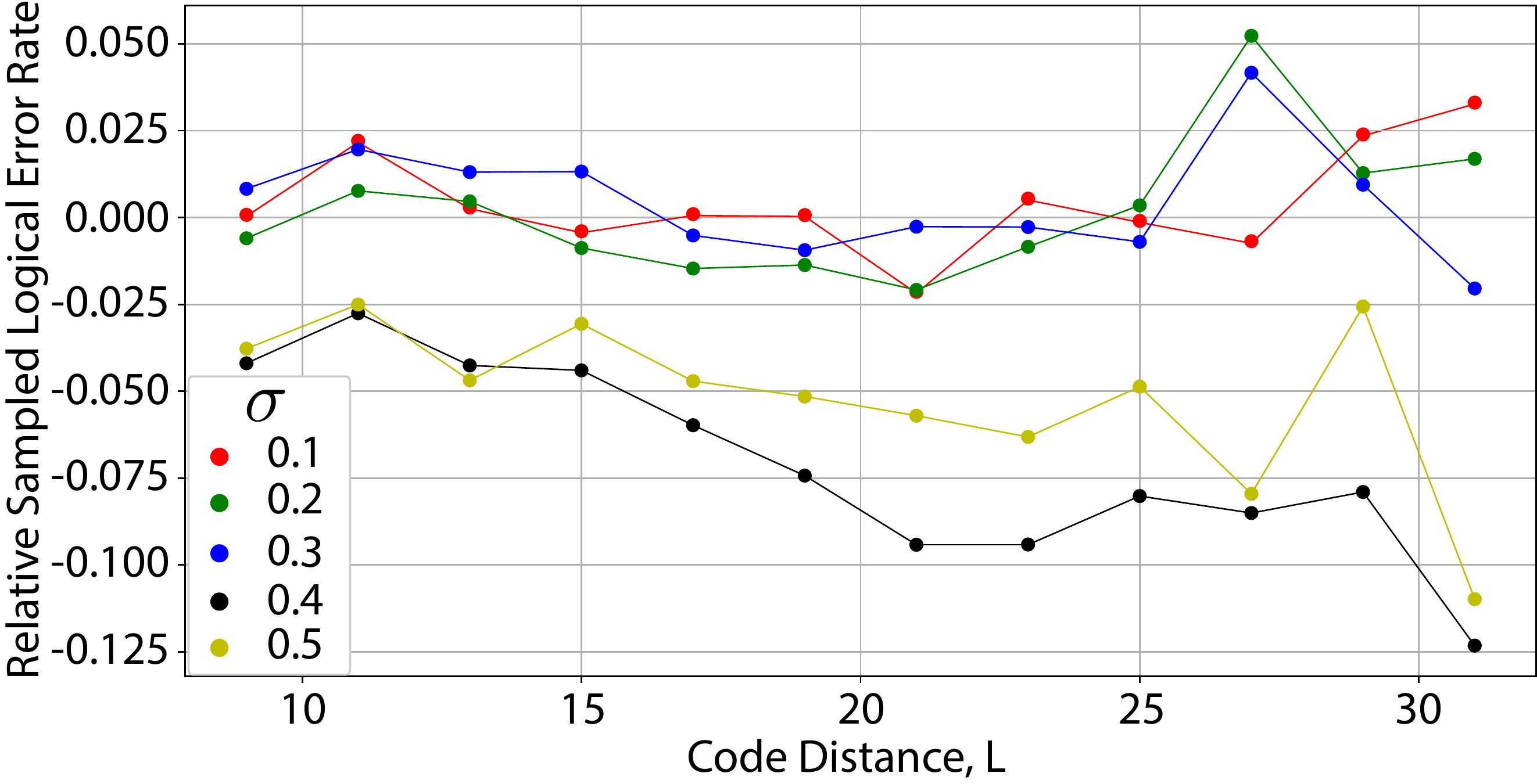}
		\caption{\label{fig:const_meas_result} Relative error rates for the repetition code under time-constant, but space-variable measurement error { as described in Section~\ref{sub:time_constant_error_rates}}. $\sigma\in\{0.1,0.2,..,0.5\}$ (red, green, blue, black, yellow) and $L\in\{9,11,..,31\}$. { $p_{\mu}=0.07$ was the physical error rate as well as the mean measurement error rate. $10^{6}$ trials were used per point.}}
	\end{center}
\end{figure}

As a simple demonstration, we take the repetition code with imperfect measurements under an error model equivalent to the phenomenological error model of the surface code \cite{Wang03a,Stephens14a}. The minimum distance between two points is found for this case by taking the horizontal Manhattan distance \cite{Manhattan19a} according to site indices, and then performing a minimisation over vertical-edge weights in the horizontal region bounded by the two points for the corresponding vertical Manhattan distance. This is in contrast to later sections where variation in time makes a full evaluation of the minimum distance across the lattice necessary. A single boolean variable to represent the parity of a qubit at the far left edge of the lattice is maintained, as is a row of 2-bit values to track the evolution of parity measurement outcomes, with the indices of odd parity outcomes passed to an extensible list in an online fashion. Correction is performed only on the tracked left-hand qubit, and is determined as the parity of the number of connections between internal lattice sites and the left lattice boundary. Assuming a final round of perfect measurement to close a single trial, the final state of this tracked qubit then records the presence or absence of logical error. Results for $10^{6}$ trials, for $\sigma\in\{0.1,..,0.5\}$, $L\in\{9,..,31\}$ and {$p_{\mu}=0.07$} are displayed in Figure~\ref{fig:const_meas_result}. We find that improvements on the order of $10$\% are observed for relative widths of order $0.4$--$0.5$, and that these improvements appear to be increasing with code distance \footnote{For the larger code distances there is some variability unexplained by the sample size. This behaviour is reproducible; we conjecture that it arises from the breakdown between error chain and error class probabilities, though the question remains open.}.

\subsection{The Impact of Local Variance}
\label{ssub:the_impact_of_local_variance}

\begin{figure}
  \centering
	\includegraphics[scale=0.4]{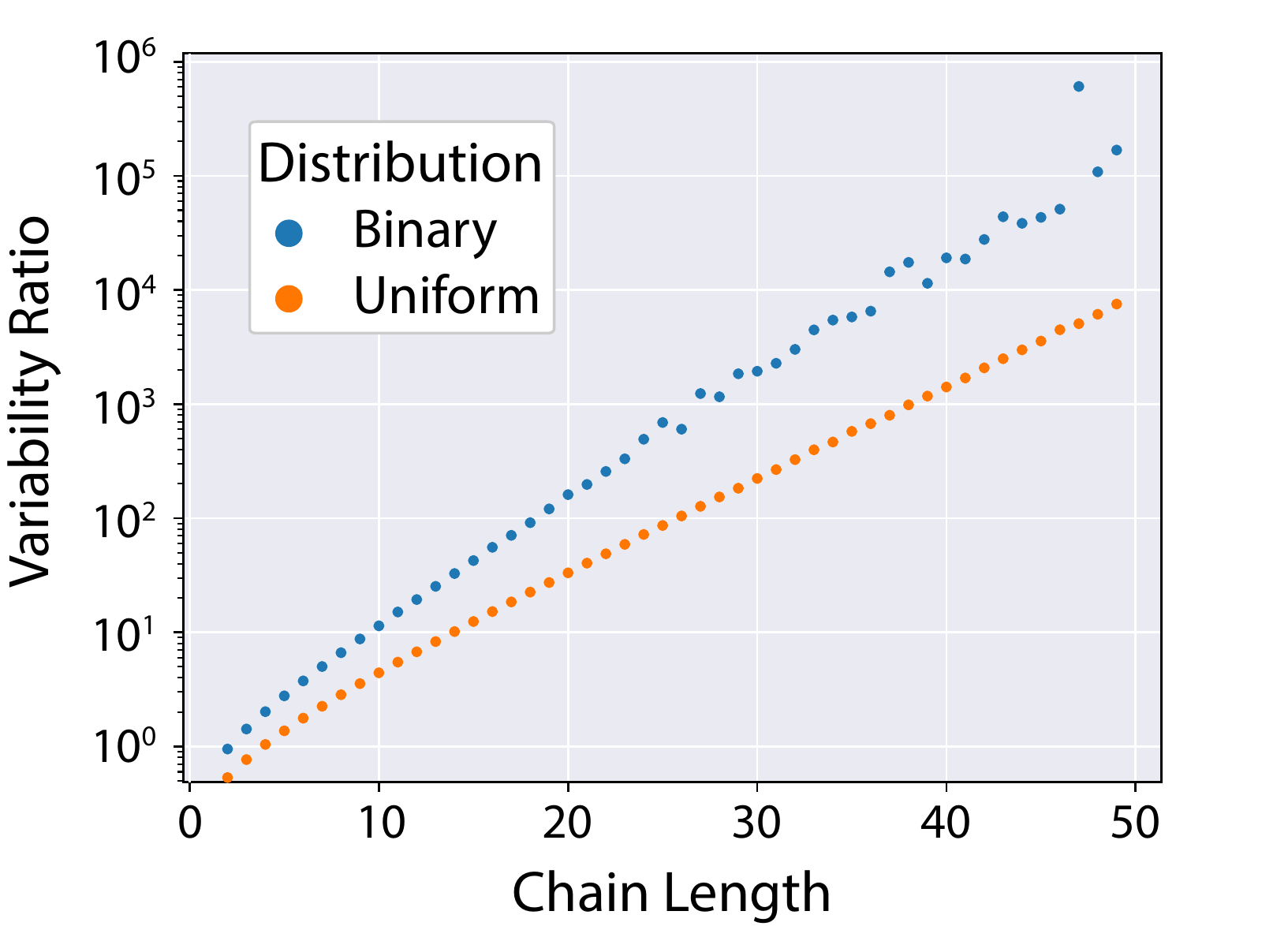}
	\caption[Ratio between the Effects of Variability and Increased Length]{The ratio between the standard deviation of a product distribution associated with a chain of length $L$ and the mean difference in probabilities between chains of lengths $L$ and $L+1$, plotted as a function of the length $L$. $10^{7}$ samples are taken for each point, and increasing variability among samples manifests in visibly increasing uncertainty in the ratio as the lengths increase. The ratio is found to increase exponentially with the lengths of the chains, and may therefore alter corrections inferred from comparisons between error chains of differing length, even with larger code distances for which the fraction $R$ of Figure~\ref{fig:noiseIntro_lOverTwoRatio} declines. \textbf{(Blue)} A discrete, balanced, two-component distribution ($\delta(x-a)/2 + \delta(x-b)/2$) with parameters $a=0.05$ and $b=0.15$. \textbf{(Orange)} A uniform distribution ($1/(b-a)$) with the same parameters.}
  \label{fig:noiseIntro_productStdDevs}
\end{figure}

\begin{figure}
  \centering
	\includegraphics[scale=0.3]{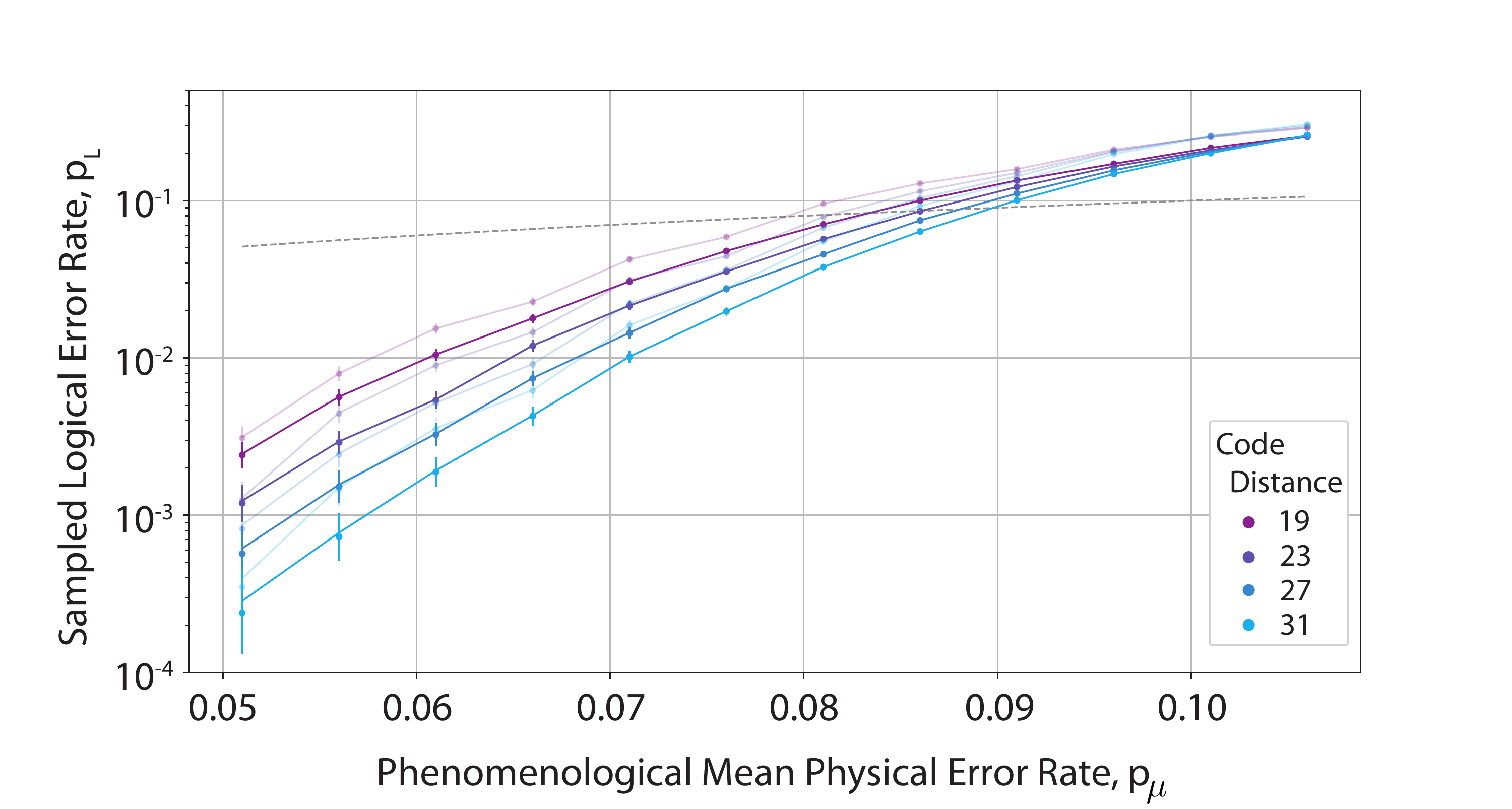}\\
  \includegraphics[scale=0.3]{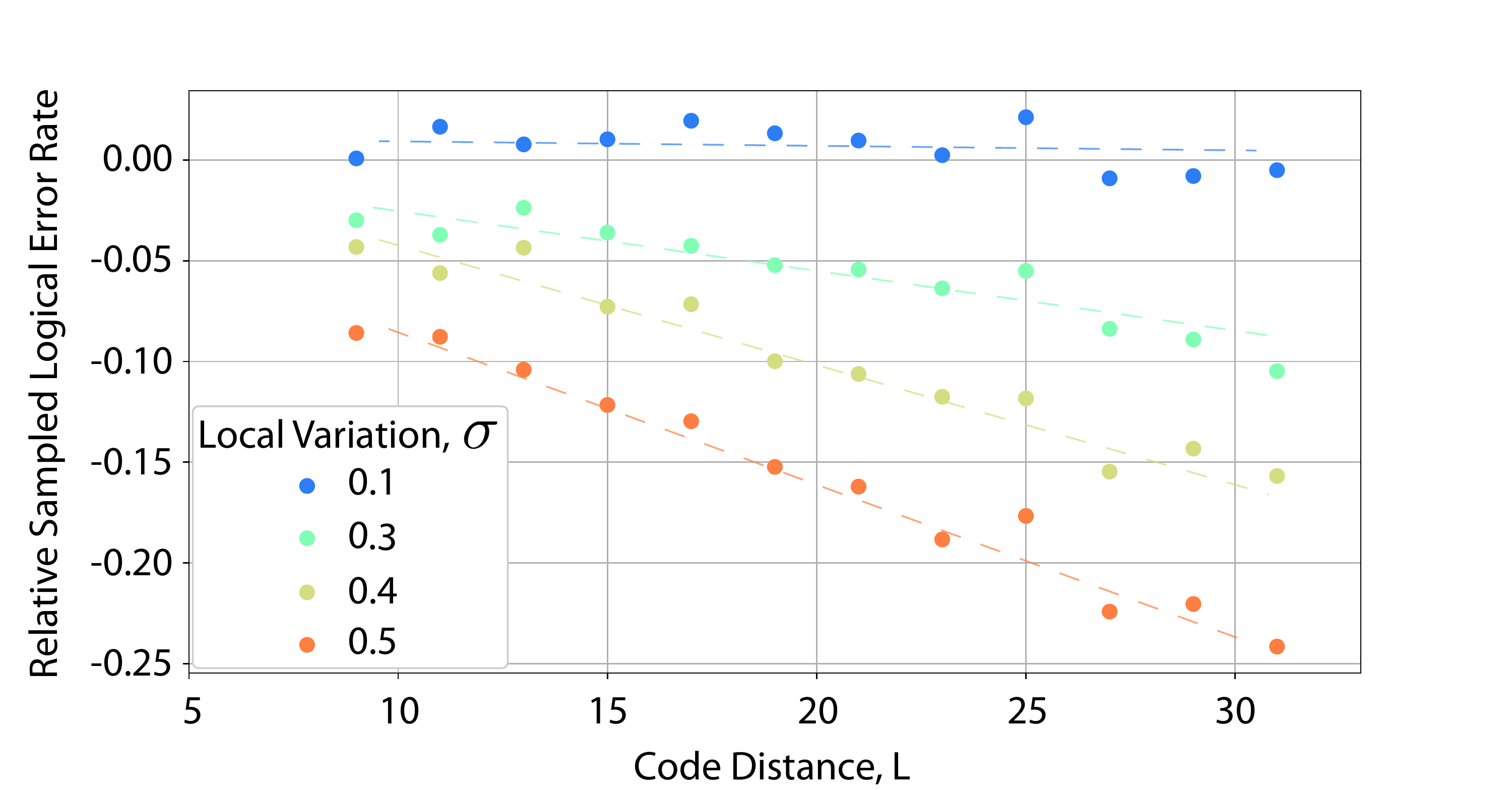}
	\caption[Repetition Code Logical Error Rates]{\textbf{(Top)} Sampled logical error rates for the repetition code over the code distances $9$, $23$, $27$, and $31$, as a function of the mean physical error rate $p_{\mu}$. Two sets of series are shown: the error rates when the mean error probability $p_{\mu}$ is used for decoding (light) and those when local variation is incorporated (dark). For the latter series, the relative local width $\sigma$ is $0.5$. Also shown is the line of equality between the two axes (grey, dashed). Error bars denote $3$ standard deviations from the mean, calculated according to the \emph{Wilson Score} \cite{Wilson27a}. \textbf{(Bottom)} \emph{Relative} change in the logical error rates when local information is incorporated, at a mean error rate $p_{\mu}=0.091$, as a function of code distances between $9$ and $31$ and for relative widths of local variation $\sigma$ between $0.1$ and $0.5$. Dashed lines are added to guide the eye. Standard estimates of sample error do not apply to this ratio distribution, but it is derived from points similar to those in the top section of this figure, for which error bars are shown. Each point in either graph is the result of $10^{5}$ trials, decoded using Kolmogorov's \emph{Blossom V} algorithm \cite{Kolmogorov09a,Edmonds65a} for minimum-weight perfect matching.}
  \label{fig:noiseIntro_rep_code_logical_error_rates}
\end{figure}

Let us now move away from the time-constant case to more general local error rates. The variance in the total weight of a sampled error chain, as its length increases, depends upon the assumed local distribution. It is not the absolute variance that is important, since this will be suppressed for longer chains, but rather the variance relative to the chains' mean weight; the variance must offset the effect of the additional multiplicative factor associated with incrementing the length of the chain. For the approximate measure defined in Section~\ref{sec:quantifying_significance}, we will look at two distributions: the uniform distribution,
\begin{align}
	D_{U}(x\in [a,b];a,b) &= \frac{1}{b-a},\label{eq:uniformDistributionDefinition}
\end{align}
and the discrete, balanced, two-component distribution,
\begin{align}
	D(x;a,b) &= \frac{\delta(x-a) + \delta(x-b)}{2}.\label{eq:binaryDistributionDefinition}
\end{align}
We calculate the ratio between the standard deviation of the weight of a chain of length $L/2$ and the difference between the mean probabilities of chains of lengths $L/2$ and $L/2 + 1$. The resultant product distributions are not normally distributed, so the standard deviation provides only a rough characterisation of the width. Without analytic formulae for the sample-product distributions, we compute the considered ratio numerically via random sampling. The results are shown in Figure~\ref{fig:noiseIntro_productStdDevs}, where the ratio is found to increase exponentially with the chain length. Local variability is therefore expected to become more significant as the code distance increases, and this is reflected in the results of our pseudo-threshold simulations, shown in Figure~\ref{fig:noiseIntro_rep_code_logical_error_rates}. This behaviour necessarily results in a slight upward shift in the accuracy threshold.

The repetition code is quite a simple code and very useful for explaining the local variance issue here. However, while the repetition code may be used for quantum sensing and in near-term biased-noise applications, we would also like to know how general this issue is for larger scale quantum computation. Let us now examine the surface code, one of the leading error correction codes being considered for large scale quantum computation \cite{Fowler12a}.

\section{The Surface Code and Chain Entropy}
\label{sub:the_surface_code}

\begin{figure}
  \centering
	\includegraphics[scale=0.08]{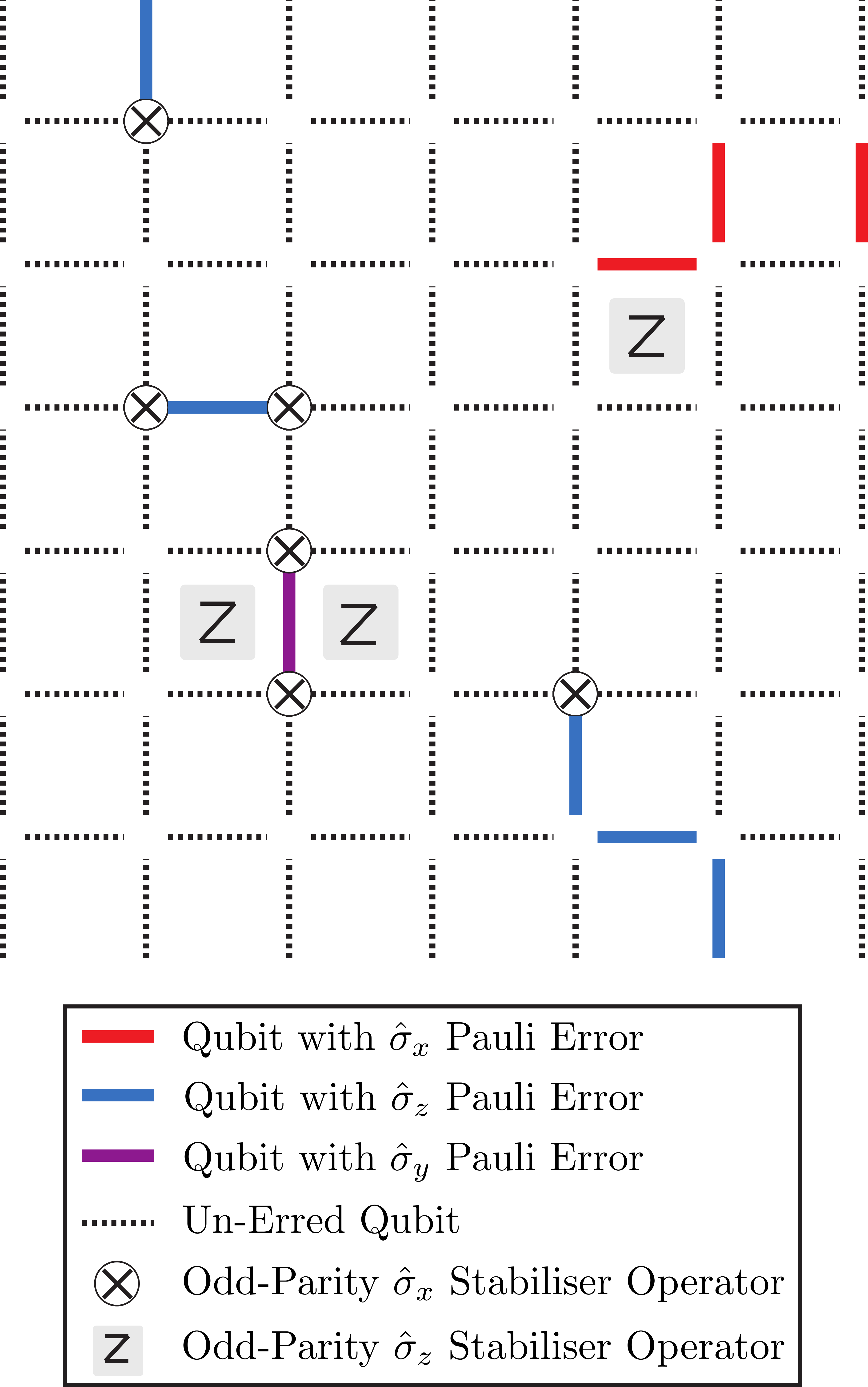}
	\caption[Surface Code Diagram]{Illustration of a length-$7$ surface code. Dotted edges correspond to qubits in their initial state. An X-basis measurement (at a vertex) detects local $\sigma_{z}$ operations and vice versa. Logical $\sigma^{(L)}_{x}$ operations stretch from the left side of the lattice to the right, while logical $\sigma^{(L)}_{z}$ operations stretch from the top edge of the lattice to the bottom. A sample error syndrome is shown; blue (red, purple) edges correspond to qubits following a local $\sigma_{z}$ ($\sigma_{x}$, $i\:\sigma_{y}$) operation.}
  \label{fig:noiseIntro_surface_code_diagram}
\end{figure}

    The surface code is defined by mapping qubits and operations to an $l\times m$ rectangular lattice \cite{Bravyi98a,Freedman01a,Dennis02a}. Edges of this lattice represent qubits, while faces and vertices represent measurements of parity operators in the $Z$ and $X$ bases respectively (the bases are arbitrary, but must be orthogonal). These measurements are defined by Pauli-operator products acting non-trivially on qubits (edges) adjacent to their respective face or vertex,
\begin{equation}
	\label{eqn:stabilizer_measurement_definition}
	\hat{S}_{Z}(f) = \prod_{e\in\partial f} \hat{\sigma}^{(e)}_{z} \quad\text{and}\quad \hat{S}_{X}(v) = \prod_{e | v\in\partial e} \hat{\sigma}^{(e)}_{x},
\end{equation}
as represented graphically in Figure~\ref{fig:noiseIntro_surface_code_diagram}. Here $v$ are vertices, $e$ are edges, $f$ are faces, and $\partial$ denotes the boundary. Vertices of degree-$1$ are excluded. The set of these measured operators generates the \emph{stabiliser} group, $S$ \cite{Gottesman97a}. Elements of this stabiliser group commute with all logical operations and therefore preserve the subspace in which the logical qubit is encoded. We require that our system exists in the $+1$ eigenspace of the stabiliser group. By then ensuring that there is exactly one more physical qubit than there are generators of this group, we restrict the total space of our system to a two-dimensional subspace within which we can define a logical qubit.

At the boundaries of the surface code, faces and vertices need not have the full complement of four adjacent edges. If a boundary consists of three-edge vertices, it is called \emph{smooth}, while if it consists of three-edge faces, it is called \emph{rough}. For the identification we have chosen, X-basis (Z-basis) operations on vertices (faces), a contiguous chain of Pauli $\sigma_{z}$ ($\sigma_{x}$) errors with both end-points at a rough (smooth) boundary will be undetectable. If both ends of the chain meet a single, contiguous such boundary, then the chain is equivalent to the application of a stabilising operation and therefore acts trivially on the logical qubit. On the other hand, if such a chain has its end-points at two non-contiguous such boundaries, then there is no equivalent stabilising operation and the chain is \emph{by definition} a logical operation. A logical operation is only unique up to elements of the stabiliser group, and is in this sense equivalent to any string of single qubit operations stretching between its two boundaries, though canonical representatives are usually defined as
\begin{align}
	\hat{\sigma}^{(L)}_{x} &= \left( \prod^{l}_{i=1} \sigma^{(i,1)}_{x} \right) \text{ and } \hat{\sigma}^{(L)}_{z} = \left( \prod^{m}_{j=1} \sigma^{(1,j)}_{z} \right),
\end{align}
where the qubits are designated on the lattice by the two dimensional indices $(i,j)$.

A surface code on an $L\times L$ lattice is a $[[L^{2}+(L-1)^{2},1,L]]$ code: it requires $L^{2}+(L-1)^{2}$ physical qubits, encodes at most $1$ logical qubit, and has a \emph{code distance} of $L$. The code distance indicates that states in the code space are topologically separated by $L$ local qubit operations. As in Section~\ref{sub:the_repetition_code}, parity measurements are repeated to account for faulty measurements, extending the code into a third \emph{time} dimension.

\subsection{Lattice Dimension}
\label{ssub:lattice_dimension}

The discussion in Section~\ref{sub:the_repetition_code} assumed a simple, one-dimensional repetition code. With the surface code as a point of comparison, we can now discuss the effect of local variation in higher dimensions. Since we are assuming that physical errors occur at a constant rate, the impact of variation will be affected by the fraction of links in a given error chain corresponding to measurement error. Increasing the dimension of the code will decrease this fraction. However, extending the lattice along an additional dimension also increases the number of qubits as well as the multiplicity of equivalent error chains: the effect of moving to higher dimensions is not trivially apparent.

The number of direct paths connecting two vertices in an $m$-dimensional lattice, when these points are separated by an equal number of links, $n$, along each dimension, is $(mn)!/(n!)^{m}$. More generally, when points are separated by a number of steps $d_{i}$ in dimension $i$, the number of direct paths is $\left( \sum_{i} d_{i} \right)! / \prod_{i} \left( d_{i}! \right) $; here we assume an average symmetry on the grounds of equal mean error rates $p_{\mu}$. The assumption that the most likely error \emph{chain} can be used as a proxy for the most likely error \emph{class} relies on the condition that the exponential suppression in likelihood with length overcomes the additional entropic contribution from the increase in the number of chains. Approximating this as
\begin{align}
	\label{eq:entropySuppression}
	\frac{(mn)!}{(n!)^{m}} p^{mn} \geq \frac{(m(n+1))!}{((n+1)!)^{m}} p^{m(n+1)},
\end{align}
and using Stirling's approximation $n!\approx \sqrt{2\pi n} (\frac{n}{e})^{n}$ we find that $p$ must satisfy
\begin{align}
	p &\leq \frac{1}{m} \left(\frac{n+1}{n}\right)^{\frac{m-1}{2m}}.
\end{align}
This approaches the finite value $p_{\text{Critical}}=1/m$ as $n$ increases, and remains larger than the accuracy thresholds of the surface code variants known at low dimensions.

\begin{figure}
  \centering
  \includegraphics[scale=0.3]{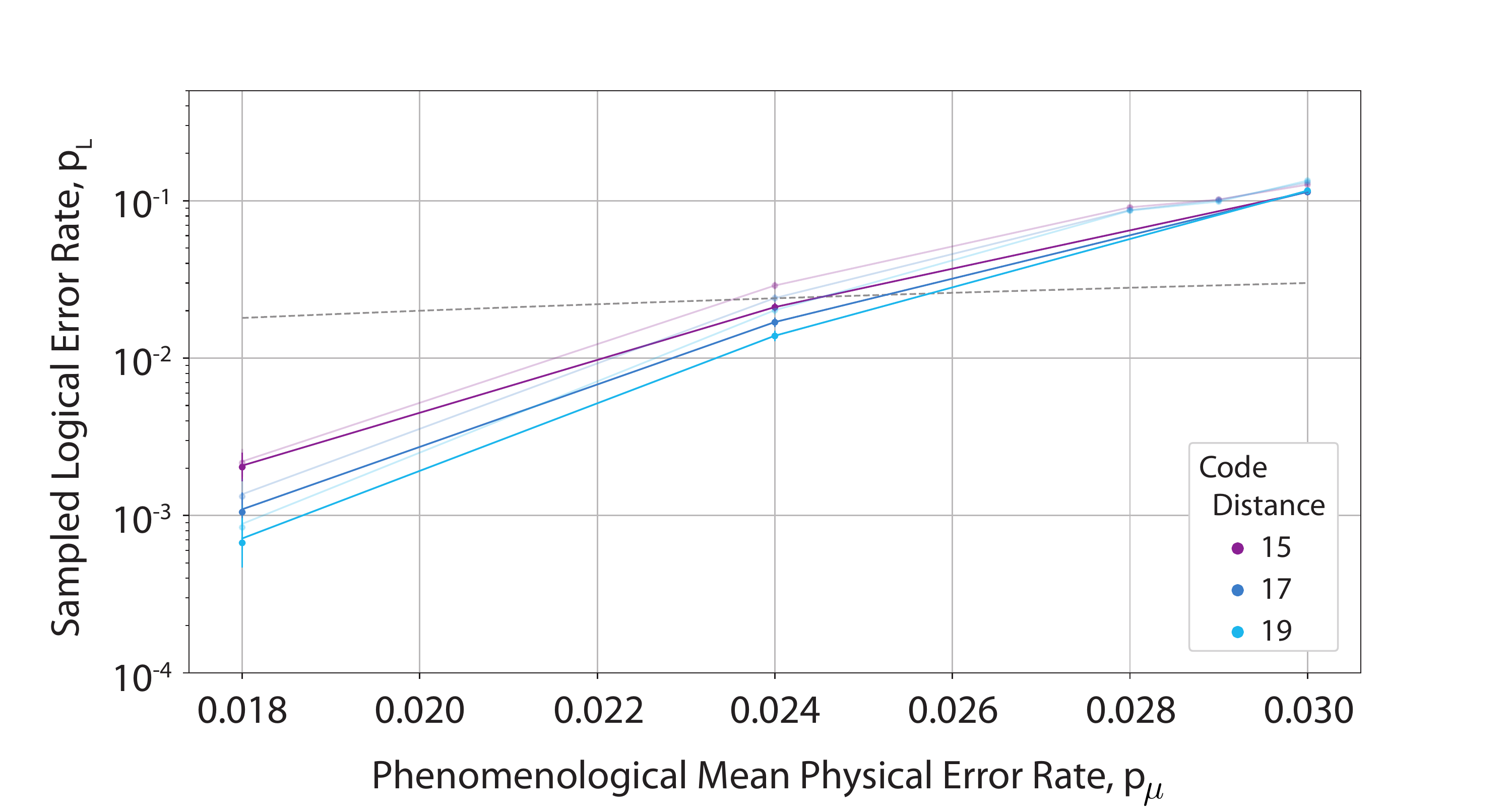}\\
  \includegraphics[scale=0.3]{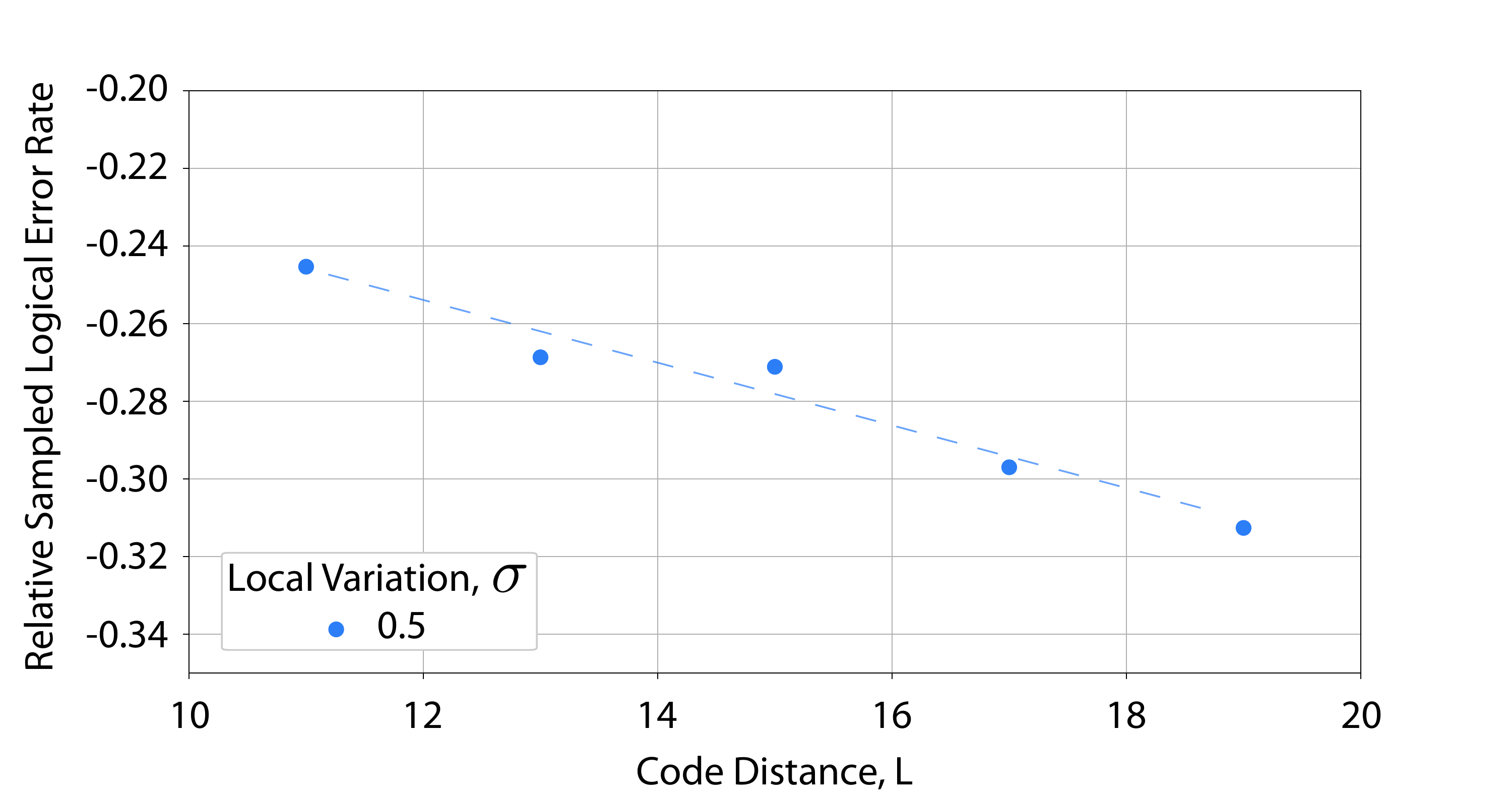}
	\caption[Surface Code Logical Error Rates]{\textbf{(Top)} Sampled logical error rates for the surface code over the code distances $15$, $17$, and $19$, as a function of the mean physical error rate $p_{\mu}$. Two sets of series are shown: the error rates when the mean error probability $p_{\mu}$ is used for decoding (light) and those when local variation is incorporated (dark). For the latter series, the relative local width $\sigma$ is $0.5$. Also shown is the line of equality between the two axes (grey, dashed). Error bars denote $3$ standard deviations from the mean, calculated according to the \emph{Wilson Score} \cite{Wilson27a}. \textbf{(Bottom)} \emph{Relative} change in the logical error rates when local information is incorporated, at a mean error rate $p_{\mu}=0.024$, as a function of code distances between $11$ and $19$ and for a relative width of local variation $\sigma = 0.5$. The dashed line is added to guide the eye. Standard estimates of sample error do not apply to this ratio distribution, but it is derived from points similar to those in the top section of this figure, for which error bars are shown. Each point in either graph is the result of $10^{5}$ trials, decoded using Kolmogorov's \emph{Blossom V} algorithm \cite{Kolmogorov09a,Edmonds65a} for minimum-weight perfect matching.}
  \label{fig:noiseIntro_code_logical_error_rates}
\end{figure}

The accuracy threshold of the code $p_{th}$ indicates the regime in which it is likely to operate, at least in the near-term. Taking the ratio between the accuracy threshold of the code $p_{th}$ and this critical probability $p_{\text{Critical}}$ gives us a measure indicating the relative impact of the diminishing probability of a chain as against the increasing multiplicity of its class. For the repetition code of Section~\ref{sub:the_repetition_code} we find that $p_{th}/p_{\text{Critical}}\approx 0.2$, while for the surface code we have $p_{th}/p_{\text{Critical}}\approx 0.09$. Here lower values are more significant. We conjecture that, given the dominance of direct error paths as indicated by our discussion in Section~\ref{sec:quantifying_significance}, the approximate factor of $2$ separating these ratios represents the relative significance of variation in the probability of a single error chain; this would be consistent with the approximate factor of $2$ between the relative improvements found for the sampled results shown in Figures~\ref{fig:noiseIntro_rep_code_logical_error_rates}~and~\ref{fig:noiseIntro_code_logical_error_rates}. To state this another way, while the accuracy threshold {is related to} a distance from the point of zero-returns, we believe this critical probability indicates a gradient {of logical error with code distance}.

\section{Discussion}
\label{sub:discussion}

In this work we have performed pseudo-threshold simulations using minimum-weight perfect matching and Kolmogorov's \emph{Blossom V} algorithm \cite{Kolmogorov09a,Edmonds65a}, and have introduced two intuitive but approximate measures of qualitative, predictive utility. Our results show that accounting for local variability in measurement errors can reduce logical error rates by factors of order $30\%$, and also show evidence that this reduction increases for higher code distances and dimensions, under the minimum-weight perfect matching decoder. There are two intuitive explanations for this behaviour: firstly, it is known that the performance of these codes under loss (which can be modelled as perfect mixing of a known subset of qubits) exceeds their performance under the depolarising channel, and we therefore expect some advantage in the spectrum between these two extremes; secondly, the gradient of the curve for logical error rate is not constant, and therefore it is not surprising that we achieve some advantage by spreading the physical error rates across this curve. In light of these points, it is not the presence but the \emph{magnitude} of the observed advantage to which we would like to draw the reader's attention.

The minimum-weight perfect matching decoder may run into difficulties at higher code distances when the weight of each chain is allowed to vary. The increase in the variance relative to the weight of the chain that we observed in Figure~\ref{fig:noiseIntro_productStdDevs} indicates that the most likely single chain becomes less representative of its entire class as the length increases. At the same time, the variance of the entire class will itself increase; individual error chains will become less significant but variability in the set of such chains should become more useful. However, the number of chains in a set increases rapidly with the length --- see Equation~\eqref{eq:entropySuppression}; when the distance between syndrome points is large, the inefficiency in classical processing required to account for the full class becomes prohibitive. An \emph{online} treatment of local variability therefore seems applicable only in small- to mid-level applications.

Beyond small- to mid-level codes, the computational cost of the minimum-weight perfect matching decoder motivates the use of alternative decoding schemes. {Belief propagation \cite{Poulin08a} would be one approach to incorporate local variation without having to consider the global lattice. By contrast, the renormalisation group decoder of Cianci et al. \cite{Cianci10a,Cianci10b,Cianci14a} is another popular alternative but} relies on pre-computed local tables. While this decoder is important because it allows us to parallelise the classical processing involved in decoding, it does not allow real-time local feedback. However, as the probability over the links of an error chain is multiplicative, the appropriate mean is geometric: the mean probability of a chain of fixed length should \emph{decrease} as the variance of individual links increases. {For some more involved error interdependence, corresponding maps from length to probability can be imagined.} The prior distribution, through its variance, therefore has a direct macroscopic impact on the logical error rate and can be accounted for even in alternative decoding schemes using pre-computed tables. Additionally, we could consider qubit rotations subject to random analogue rotation errors, without the local feedback from measurement.

Finally, we note that the impact of measurement error is magnified for measurement-based quantum computing, for which it is involved in every operation, and for distributed schemes, wherein heralded transmission losses allow a distinct source of local information about operation error rates. We expect local variability to provide large effective reductions in resource requirements in the near-term, as resources are severely limited and gate error rates for many systems remain at or near the surface code accuracy threshold of $\sim 1\%$.

\begin{acknowledgements}
We thank Simon Devitt and Ashley Stephens for useful discussions. { Further, Ashley Stephens was kind enough to help us confirm a suspicion about some early results.} This work was supported in part from the Japanese program MEXT Q-LEAP (Funding No. JPMXS0118069605), the MEXT KAKENHI Grant-in-Aid for Scientific Research on Innovative Areas Science of hybrid quantum systems Grant No.15H05870 and the JSPS KAKENHI Grant No. 19H00662. This publication was made possible through the support of a grant from the John Templeton Foundation (JTF \# 60478). The opinions expressed in this publication are those of the author(s) and do not necessarily reflect the views of the John Templeton Foundation.
\end{acknowledgements}

\bibliography{bibliography}

\begin{thebibliography}{57}%
\makeatletter
\providecommand \@ifxundefined [1]{%
 \@ifx{#1\undefined}
}%
\providecommand \@ifnum [1]{%
 \ifnum #1\expandafter \@firstoftwo
 \else \expandafter \@secondoftwo
 \fi
}%
\providecommand \@ifx [1]{%
 \ifx #1\expandafter \@firstoftwo
 \else \expandafter \@secondoftwo
 \fi
}%
\providecommand \natexlab [1]{#1}%
\providecommand \enquote  [1]{``#1''}%
\providecommand \bibnamefont  [1]{#1}%
\providecommand \bibfnamefont [1]{#1}%
\providecommand \citenamefont [1]{#1}%
\providecommand \href@noop [0]{\@secondoftwo}%
\providecommand \href [0]{\begingroup \@sanitize@url \@href}%
\providecommand \@href[1]{\@@startlink{#1}\@@href}%
\providecommand \@@href[1]{\endgroup#1\@@endlink}%
\providecommand \@sanitize@url [0]{\catcode `\\12\catcode `\$12\catcode
  `\&12\catcode `\#12\catcode `\^12\catcode `\_12\catcode `\%12\relax}%
\providecommand \@@startlink[1]{}%
\providecommand \@@endlink[0]{}%
\providecommand \url  [0]{\begingroup\@sanitize@url \@url }%
\providecommand \@url [1]{\endgroup\@href {#1}{\urlprefix }}%
\providecommand \urlprefix  [0]{URL }%
\providecommand \Eprint [0]{\href }%
\providecommand \doibase [0]{http://dx.doi.org/}%
\providecommand \selectlanguage [0]{\@gobble}%
\providecommand \bibinfo  [0]{\@secondoftwo}%
\providecommand \bibfield  [0]{\@secondoftwo}%
\providecommand \translation [1]{[#1]}%
\providecommand \BibitemOpen [0]{}%
\providecommand \bibitemStop [0]{}%
\providecommand \bibitemNoStop [0]{.\EOS\space}%
\providecommand \EOS [0]{\spacefactor3000\relax}%
\providecommand \BibitemShut  [1]{\csname bibitem#1\endcsname}%
\let\auto@bib@innerbib\@empty
\bibitem [{\citenamefont {Landauer}(1995)}]{Landauer95a}%
  \BibitemOpen
  \bibfield  {author} {\bibinfo {author} {\bibfnamefont {R.}~\bibnamefont
  {Landauer}},\ }\href@noop {} {\bibfield  {journal} {\bibinfo  {journal}
  {Philosophical Transactions of the Royal Society of London. Series A:
  Physical and Engineering Sciences}\ }\textbf {\bibinfo {volume} {353}},\
  \bibinfo {pages} {367} (\bibinfo {year} {1995})}\BibitemShut {NoStop}%
\bibitem [{\citenamefont {Chuang}\ \emph {et~al.}(1995)\citenamefont {Chuang},
  \citenamefont {Laflamme}, \citenamefont {Shor},\ and\ \citenamefont
  {Zurek}}]{Chuang95a}%
  \BibitemOpen
  \bibfield  {author} {\bibinfo {author} {\bibfnamefont {I.~L.}\ \bibnamefont
  {Chuang}}, \bibinfo {author} {\bibfnamefont {R.}~\bibnamefont {Laflamme}},
  \bibinfo {author} {\bibfnamefont {P.~W.}\ \bibnamefont {Shor}}, \ and\
  \bibinfo {author} {\bibfnamefont {W.~H.}\ \bibnamefont {Zurek}},\ }\href@noop
  {} {\bibfield  {journal} {\bibinfo  {journal} {Science}\ }\textbf {\bibinfo
  {volume} {270}},\ \bibinfo {pages} {1633} (\bibinfo {year}
  {1995})}\BibitemShut {NoStop}%
\bibitem [{\citenamefont {Unruh}(1995)}]{Unruh95a}%
  \BibitemOpen
  \bibfield  {author} {\bibinfo {author} {\bibfnamefont {W.~G.}\ \bibnamefont
  {Unruh}},\ }\href@noop {} {\bibfield  {journal} {\bibinfo  {journal}
  {Physical Review A}\ }\textbf {\bibinfo {volume} {51}},\ \bibinfo {pages}
  {992} (\bibinfo {year} {1995})}\BibitemShut {NoStop}%
\bibitem [{\citenamefont {Palma}\ \emph {et~al.}(1996)\citenamefont {Palma},
  \citenamefont {Suominen},\ and\ \citenamefont {Ekert}}]{Palma96a}%
  \BibitemOpen
  \bibfield  {author} {\bibinfo {author} {\bibfnamefont {G.~M.}\ \bibnamefont
  {Palma}}, \bibinfo {author} {\bibfnamefont {K.-A.}\ \bibnamefont {Suominen}},
  \ and\ \bibinfo {author} {\bibfnamefont {A.}~\bibnamefont {Ekert}},\
  }\href@noop {} {\bibfield  {journal} {\bibinfo  {journal} {Proceedings of the
  Royal Society of London. Series A: Mathematical, Physical and Engineering
  Sciences}\ }\textbf {\bibinfo {volume} {452}},\ \bibinfo {pages} {567}
  (\bibinfo {year} {1996})}\BibitemShut {NoStop}%
\bibitem [{\citenamefont {Lidar}\ and\ \citenamefont {Brun}(2013)}]{Lidar13a}%
  \BibitemOpen
  \bibfield  {author} {\bibinfo {author} {\bibfnamefont {D.~A.}\ \bibnamefont
  {Lidar}}\ and\ \bibinfo {author} {\bibfnamefont {T.~A.}\ \bibnamefont
  {Brun}},\ }\href@noop {} {\emph {\bibinfo {title} {Quantum error
  correction}}}\ (\bibinfo  {publisher} {Cambridge university press},\ \bibinfo
  {year} {2013})\BibitemShut {NoStop}%
\bibitem [{\citenamefont {Devitt}\ \emph {et~al.}(2013)\citenamefont {Devitt},
  \citenamefont {Munro},\ and\ \citenamefont {Nemoto}}]{Devitt13a}%
  \BibitemOpen
  \bibfield  {author} {\bibinfo {author} {\bibfnamefont {S.~J.}\ \bibnamefont
  {Devitt}}, \bibinfo {author} {\bibfnamefont {W.~J.}\ \bibnamefont {Munro}}, \
  and\ \bibinfo {author} {\bibfnamefont {K.}~\bibnamefont {Nemoto}},\
  }\href@noop {} {\bibfield  {journal} {\bibinfo  {journal} {Reports on
  Progress in Physics}\ }\textbf {\bibinfo {volume} {76}},\ \bibinfo {pages}
  {076001} (\bibinfo {year} {2013})}\BibitemShut {NoStop}%
\bibitem [{\citenamefont {Terhal}(2015)}]{Terhal15a}%
  \BibitemOpen
  \bibfield  {author} {\bibinfo {author} {\bibfnamefont {B.~M.}\ \bibnamefont
  {Terhal}},\ }\href@noop {} {\bibfield  {journal} {\bibinfo  {journal}
  {Reviews of Modern Physics}\ }\textbf {\bibinfo {volume} {87}},\ \bibinfo
  {pages} {307} (\bibinfo {year} {2015})}\BibitemShut {NoStop}%
\bibitem [{\citenamefont {Arrad}\ \emph {et~al.}(2014)\citenamefont {Arrad},
  \citenamefont {Vinkler}, \citenamefont {Aharonov},\ and\ \citenamefont
  {Retzker}}]{Arrad14a}%
  \BibitemOpen
  \bibfield  {author} {\bibinfo {author} {\bibfnamefont {G.}~\bibnamefont
  {Arrad}}, \bibinfo {author} {\bibfnamefont {Y.}~\bibnamefont {Vinkler}},
  \bibinfo {author} {\bibfnamefont {D.}~\bibnamefont {Aharonov}}, \ and\
  \bibinfo {author} {\bibfnamefont {A.}~\bibnamefont {Retzker}},\ }\href@noop
  {} {\bibfield  {journal} {\bibinfo  {journal} {Physical review letters}\
  }\textbf {\bibinfo {volume} {112}},\ \bibinfo {pages} {150801} (\bibinfo
  {year} {2014})}\BibitemShut {NoStop}%
\bibitem [{\citenamefont {Kessler}\ \emph {et~al.}(2014)\citenamefont
  {Kessler}, \citenamefont {Lovchinsky}, \citenamefont {Sushkov},\ and\
  \citenamefont {Lukin}}]{Kessler14a}%
  \BibitemOpen
  \bibfield  {author} {\bibinfo {author} {\bibfnamefont {E.~M.}\ \bibnamefont
  {Kessler}}, \bibinfo {author} {\bibfnamefont {I.}~\bibnamefont {Lovchinsky}},
  \bibinfo {author} {\bibfnamefont {A.~O.}\ \bibnamefont {Sushkov}}, \ and\
  \bibinfo {author} {\bibfnamefont {M.~D.}\ \bibnamefont {Lukin}},\ }\href@noop
  {} {\bibfield  {journal} {\bibinfo  {journal} {Physical review letters}\
  }\textbf {\bibinfo {volume} {112}},\ \bibinfo {pages} {150802} (\bibinfo
  {year} {2014})}\BibitemShut {NoStop}%
\bibitem [{\citenamefont {D{\"u}r}\ \emph {et~al.}(2014)\citenamefont
  {D{\"u}r}, \citenamefont {Skotiniotis}, \citenamefont {Froewis},\ and\
  \citenamefont {Kraus}}]{Duer14a}%
  \BibitemOpen
  \bibfield  {author} {\bibinfo {author} {\bibfnamefont {W.}~\bibnamefont
  {D{\"u}r}}, \bibinfo {author} {\bibfnamefont {M.}~\bibnamefont
  {Skotiniotis}}, \bibinfo {author} {\bibfnamefont {F.}~\bibnamefont
  {Froewis}}, \ and\ \bibinfo {author} {\bibfnamefont {B.}~\bibnamefont
  {Kraus}},\ }\href@noop {} {\bibfield  {journal} {\bibinfo  {journal}
  {Physical Review Letters}\ }\textbf {\bibinfo {volume} {112}},\ \bibinfo
  {pages} {080801} (\bibinfo {year} {2014})}\BibitemShut {NoStop}%
\bibitem [{\citenamefont {Unden}\ \emph {et~al.}(2016)\citenamefont {Unden},
  \citenamefont {Balasubramanian}, \citenamefont {Louzon}, \citenamefont
  {Vinkler}, \citenamefont {Plenio}, \citenamefont {Markham}, \citenamefont
  {Twitchen}, \citenamefont {Stacey}, \citenamefont {Lovchinsky}, \citenamefont
  {Sushkov} \emph {et~al.}}]{Unden16a}%
  \BibitemOpen
  \bibfield  {author} {\bibinfo {author} {\bibfnamefont {T.}~\bibnamefont
  {Unden}}, \bibinfo {author} {\bibfnamefont {P.}~\bibnamefont
  {Balasubramanian}}, \bibinfo {author} {\bibfnamefont {D.}~\bibnamefont
  {Louzon}}, \bibinfo {author} {\bibfnamefont {Y.}~\bibnamefont {Vinkler}},
  \bibinfo {author} {\bibfnamefont {M.~B.}\ \bibnamefont {Plenio}}, \bibinfo
  {author} {\bibfnamefont {M.}~\bibnamefont {Markham}}, \bibinfo {author}
  {\bibfnamefont {D.}~\bibnamefont {Twitchen}}, \bibinfo {author}
  {\bibfnamefont {A.}~\bibnamefont {Stacey}}, \bibinfo {author} {\bibfnamefont
  {I.}~\bibnamefont {Lovchinsky}}, \bibinfo {author} {\bibfnamefont {A.~O.}\
  \bibnamefont {Sushkov}},  \emph {et~al.},\ }\href@noop {} {\bibfield
  {journal} {\bibinfo  {journal} {Physical review letters}\ }\textbf {\bibinfo
  {volume} {116}},\ \bibinfo {pages} {230502} (\bibinfo {year}
  {2016})}\BibitemShut {NoStop}%
\bibitem [{\citenamefont {Bravyi}\ and\ \citenamefont
  {Kitaev}(1998)}]{Bravyi98a}%
  \BibitemOpen
  \bibfield  {author} {\bibinfo {author} {\bibfnamefont {S.~B.}\ \bibnamefont
  {Bravyi}}\ and\ \bibinfo {author} {\bibfnamefont {A.~Y.}\ \bibnamefont
  {Kitaev}},\ }\href@noop {} {\bibfield  {journal} {\bibinfo  {journal} {arXiv
  preprint quant-ph/9811052}\ } (\bibinfo {year} {1998})}\BibitemShut {NoStop}%
\bibitem [{\citenamefont {Freedman}\ and\ \citenamefont
  {Meyer}(2001)}]{Freedman01a}%
  \BibitemOpen
  \bibfield  {author} {\bibinfo {author} {\bibfnamefont {M.~H.}\ \bibnamefont
  {Freedman}}\ and\ \bibinfo {author} {\bibfnamefont {D.~A.}\ \bibnamefont
  {Meyer}},\ }\href@noop {} {\bibfield  {journal} {\bibinfo  {journal}
  {Foundations of Computational Mathematics}\ }\textbf {\bibinfo {volume}
  {1}},\ \bibinfo {pages} {325} (\bibinfo {year} {2001})}\BibitemShut {NoStop}%
\bibitem [{\citenamefont {Dennis}\ \emph {et~al.}(2002)\citenamefont {Dennis},
  \citenamefont {Kitaev}, \citenamefont {Landahl},\ and\ \citenamefont
  {Preskill}}]{Dennis02a}%
  \BibitemOpen
  \bibfield  {author} {\bibinfo {author} {\bibfnamefont {E.}~\bibnamefont
  {Dennis}}, \bibinfo {author} {\bibfnamefont {A.}~\bibnamefont {Kitaev}},
  \bibinfo {author} {\bibfnamefont {A.}~\bibnamefont {Landahl}}, \ and\
  \bibinfo {author} {\bibfnamefont {J.}~\bibnamefont {Preskill}},\ }\href@noop
  {} {\bibfield  {journal} {\bibinfo  {journal} {Journal of Mathematical
  Physics}\ }\textbf {\bibinfo {volume} {43}},\ \bibinfo {pages} {4452}
  (\bibinfo {year} {2002})}\BibitemShut {NoStop}%
\bibitem [{\citenamefont {Shor}(1995)}]{Shor95a}%
  \BibitemOpen
  \bibfield  {author} {\bibinfo {author} {\bibfnamefont {P.~W.}\ \bibnamefont
  {Shor}},\ }\href@noop {} {\bibfield  {journal} {\bibinfo  {journal} {Physical
  review A}\ }\textbf {\bibinfo {volume} {52}},\ \bibinfo {pages} {R2493}
  (\bibinfo {year} {1995})}\BibitemShut {NoStop}%
\bibitem [{\citenamefont {Knill}\ and\ \citenamefont
  {Laflamme}(1996)}]{Knill96a}%
  \BibitemOpen
  \bibfield  {author} {\bibinfo {author} {\bibfnamefont {E.}~\bibnamefont
  {Knill}}\ and\ \bibinfo {author} {\bibfnamefont {R.}~\bibnamefont
  {Laflamme}},\ }\href@noop {} {\bibfield  {journal} {\bibinfo  {journal}
  {arXiv preprint quant-ph/9608012}\ } (\bibinfo {year} {1996})}\BibitemShut
  {NoStop}%
\bibitem [{\citenamefont {Berthiaume}\ \emph {et~al.}(1994)\citenamefont
  {Berthiaume}, \citenamefont {Deutsch},\ and\ \citenamefont
  {Jozsa}}]{Berthiaume94a}%
  \BibitemOpen
  \bibfield  {author} {\bibinfo {author} {\bibfnamefont {A.}~\bibnamefont
  {Berthiaume}}, \bibinfo {author} {\bibfnamefont {D.}~\bibnamefont {Deutsch}},
  \ and\ \bibinfo {author} {\bibfnamefont {R.}~\bibnamefont {Jozsa}},\ }in\
  \href@noop {} {\emph {\bibinfo {booktitle} {Proceedings Workshop on Physics
  and Computation. PhysComp'94}}}\ (\bibinfo {organization} {IEEE},\ \bibinfo
  {year} {1994})\ pp.\ \bibinfo {pages} {60--62}\BibitemShut {NoStop}%
\bibitem [{\citenamefont {Bennett}\ \emph {et~al.}(1994)\citenamefont
  {Bennett}, \citenamefont {Brassard}, \citenamefont {Jozsa}, \citenamefont
  {Mayers}, \citenamefont {Peres}, \citenamefont {Schumacher},\ and\
  \citenamefont {Wootters}}]{Bennett94a}%
  \BibitemOpen
  \bibfield  {author} {\bibinfo {author} {\bibfnamefont {C.~H.}\ \bibnamefont
  {Bennett}}, \bibinfo {author} {\bibfnamefont {G.}~\bibnamefont {Brassard}},
  \bibinfo {author} {\bibfnamefont {R.}~\bibnamefont {Jozsa}}, \bibinfo
  {author} {\bibfnamefont {D.}~\bibnamefont {Mayers}}, \bibinfo {author}
  {\bibfnamefont {A.}~\bibnamefont {Peres}}, \bibinfo {author} {\bibfnamefont
  {B.}~\bibnamefont {Schumacher}}, \ and\ \bibinfo {author} {\bibfnamefont
  {W.~K.}\ \bibnamefont {Wootters}},\ }\href@noop {} {\bibfield  {journal}
  {\bibinfo  {journal} {Journal of Modern Optics}\ }\textbf {\bibinfo {volume}
  {41}},\ \bibinfo {pages} {2307} (\bibinfo {year} {1994})}\BibitemShut
  {NoStop}%
\bibitem [{\citenamefont {Vaidman}\ \emph {et~al.}(1996)\citenamefont
  {Vaidman}, \citenamefont {Goldenberg},\ and\ \citenamefont
  {Wiesner}}]{Vaidman96a}%
  \BibitemOpen
  \bibfield  {author} {\bibinfo {author} {\bibfnamefont {L.}~\bibnamefont
  {Vaidman}}, \bibinfo {author} {\bibfnamefont {L.}~\bibnamefont {Goldenberg}},
  \ and\ \bibinfo {author} {\bibfnamefont {S.}~\bibnamefont {Wiesner}},\
  }\href@noop {} {\bibfield  {journal} {\bibinfo  {journal} {Physical Review
  A}\ }\textbf {\bibinfo {volume} {54}},\ \bibinfo {pages} {R1745} (\bibinfo
  {year} {1996})}\BibitemShut {NoStop}%
\bibitem [{\citenamefont {Barenco}\ \emph {et~al.}(1997)\citenamefont
  {Barenco}, \citenamefont {Berthiaume}, \citenamefont {Deutsch}, \citenamefont
  {Ekert}, \citenamefont {Jozsa},\ and\ \citenamefont
  {Macchiavello}}]{Barenco97a}%
  \BibitemOpen
  \bibfield  {author} {\bibinfo {author} {\bibfnamefont {A.}~\bibnamefont
  {Barenco}}, \bibinfo {author} {\bibfnamefont {A.}~\bibnamefont {Berthiaume}},
  \bibinfo {author} {\bibfnamefont {D.}~\bibnamefont {Deutsch}}, \bibinfo
  {author} {\bibfnamefont {A.}~\bibnamefont {Ekert}}, \bibinfo {author}
  {\bibfnamefont {R.}~\bibnamefont {Jozsa}}, \ and\ \bibinfo {author}
  {\bibfnamefont {C.}~\bibnamefont {Macchiavello}},\ }\href@noop {} {\bibfield
  {journal} {\bibinfo  {journal} {SIAM Journal on Computing}\ }\textbf
  {\bibinfo {volume} {26}},\ \bibinfo {pages} {1541} (\bibinfo {year}
  {1997})}\BibitemShut {NoStop}%
\bibitem [{\citenamefont {Misra}\ and\ \citenamefont
  {Sudarshan}(1977)}]{Misra77a}%
  \BibitemOpen
  \bibfield  {author} {\bibinfo {author} {\bibfnamefont {B.}~\bibnamefont
  {Misra}}\ and\ \bibinfo {author} {\bibfnamefont {E.~G.}\ \bibnamefont
  {Sudarshan}},\ }\href@noop {} {\bibfield  {journal} {\bibinfo  {journal}
  {Journal of Mathematical Physics}\ }\textbf {\bibinfo {volume} {18}},\
  \bibinfo {pages} {756} (\bibinfo {year} {1977})}\BibitemShut {NoStop}%
\bibitem [{\citenamefont {Bennett}\ \emph {et~al.}(1996)\citenamefont
  {Bennett}, \citenamefont {DiVincenzo}, \citenamefont {Smolin},\ and\
  \citenamefont {Wootters}}]{Bennett96a}%
  \BibitemOpen
  \bibfield  {author} {\bibinfo {author} {\bibfnamefont {C.~H.}\ \bibnamefont
  {Bennett}}, \bibinfo {author} {\bibfnamefont {D.~P.}\ \bibnamefont
  {DiVincenzo}}, \bibinfo {author} {\bibfnamefont {J.~A.}\ \bibnamefont
  {Smolin}}, \ and\ \bibinfo {author} {\bibfnamefont {W.~K.}\ \bibnamefont
  {Wootters}},\ }\href@noop {} {\bibfield  {journal} {\bibinfo  {journal}
  {Physical Review A}\ }\textbf {\bibinfo {volume} {54}},\ \bibinfo {pages}
  {3824} (\bibinfo {year} {1996})}\BibitemShut {NoStop}%
\bibitem [{\citenamefont {Ralph}\ \emph {et~al.}(2005)\citenamefont {Ralph},
  \citenamefont {Hayes},\ and\ \citenamefont {Gilchrist}}]{Ralph05a}%
  \BibitemOpen
  \bibfield  {author} {\bibinfo {author} {\bibfnamefont {T.~C.}\ \bibnamefont
  {Ralph}}, \bibinfo {author} {\bibfnamefont {A.}~\bibnamefont {Hayes}}, \ and\
  \bibinfo {author} {\bibfnamefont {A.}~\bibnamefont {Gilchrist}},\ }\href@noop
  {} {\bibfield  {journal} {\bibinfo  {journal} {Physical review letters}\
  }\textbf {\bibinfo {volume} {95}},\ \bibinfo {pages} {100501} (\bibinfo
  {year} {2005})}\BibitemShut {NoStop}%
\bibitem [{\citenamefont {Aliferis}\ and\ \citenamefont
  {Preskill}(2008)}]{Aliferis08a}%
  \BibitemOpen
  \bibfield  {author} {\bibinfo {author} {\bibfnamefont {P.}~\bibnamefont
  {Aliferis}}\ and\ \bibinfo {author} {\bibfnamefont {J.}~\bibnamefont
  {Preskill}},\ }\href@noop {} {\bibfield  {journal} {\bibinfo  {journal}
  {Physical Review A}\ }\textbf {\bibinfo {volume} {78}},\ \bibinfo {pages}
  {052331} (\bibinfo {year} {2008})}\BibitemShut {NoStop}%
\bibitem [{\citenamefont {Brooks}(2013)}]{Brooks13a}%
  \BibitemOpen
  \bibfield  {author} {\bibinfo {author} {\bibfnamefont {P.~B.}\ \bibnamefont
  {Brooks}},\ }\emph {\bibinfo {title} {Quantum error correction with biased
  noise}},\ \href@noop {} {Ph.D. thesis},\ \bibinfo  {school} {California
  Institute of Technology} (\bibinfo {year} {2013})\BibitemShut {NoStop}%
\bibitem [{\citenamefont {Stephens}\ \emph {et~al.}(2013)\citenamefont
  {Stephens}, \citenamefont {Munro},\ and\ \citenamefont
  {Nemoto}}]{Stephens13a}%
  \BibitemOpen
  \bibfield  {author} {\bibinfo {author} {\bibfnamefont {A.~M.}\ \bibnamefont
  {Stephens}}, \bibinfo {author} {\bibfnamefont {W.~J.}\ \bibnamefont {Munro}},
  \ and\ \bibinfo {author} {\bibfnamefont {K.}~\bibnamefont {Nemoto}},\
  }\href@noop {} {\bibfield  {journal} {\bibinfo  {journal} {Physical Review
  A}\ }\textbf {\bibinfo {volume} {88}},\ \bibinfo {pages} {060301} (\bibinfo
  {year} {2013})}\BibitemShut {NoStop}%
\bibitem [{\citenamefont {Stace}\ \emph {et~al.}(2009)\citenamefont {Stace},
  \citenamefont {Barrett},\ and\ \citenamefont {Doherty}}]{Stace09a}%
  \BibitemOpen
  \bibfield  {author} {\bibinfo {author} {\bibfnamefont {T.~M.}\ \bibnamefont
  {Stace}}, \bibinfo {author} {\bibfnamefont {S.~D.}\ \bibnamefont {Barrett}},
  \ and\ \bibinfo {author} {\bibfnamefont {A.~C.}\ \bibnamefont {Doherty}},\
  }\href@noop {} {\bibfield  {journal} {\bibinfo  {journal} {Physical review
  letters}\ }\textbf {\bibinfo {volume} {102}},\ \bibinfo {pages} {200501}
  (\bibinfo {year} {2009})}\BibitemShut {NoStop}%
\bibitem [{\citenamefont {Stace}\ and\ \citenamefont
  {Barrett}(2010)}]{Stace10a}%
  \BibitemOpen
  \bibfield  {author} {\bibinfo {author} {\bibfnamefont {T.~M.}\ \bibnamefont
  {Stace}}\ and\ \bibinfo {author} {\bibfnamefont {S.~D.}\ \bibnamefont
  {Barrett}},\ }\href@noop {} {\bibfield  {journal} {\bibinfo  {journal}
  {Physical Review A}\ }\textbf {\bibinfo {volume} {81}},\ \bibinfo {pages}
  {022317} (\bibinfo {year} {2010})}\BibitemShut {NoStop}%
\bibitem [{\citenamefont {Barrett}\ and\ \citenamefont
  {Stace}(2010)}]{Barrett10a}%
  \BibitemOpen
  \bibfield  {author} {\bibinfo {author} {\bibfnamefont {S.~D.}\ \bibnamefont
  {Barrett}}\ and\ \bibinfo {author} {\bibfnamefont {T.~M.}\ \bibnamefont
  {Stace}},\ }\href@noop {} {\bibfield  {journal} {\bibinfo  {journal}
  {Physical review letters}\ }\textbf {\bibinfo {volume} {105}},\ \bibinfo
  {pages} {200502} (\bibinfo {year} {2010})}\BibitemShut {NoStop}%
\bibitem [{\citenamefont {Ohzeki}(2012)}]{Ohzeki12a}%
  \BibitemOpen
  \bibfield  {author} {\bibinfo {author} {\bibfnamefont {M.}~\bibnamefont
  {Ohzeki}},\ }\href@noop {} {\bibfield  {journal} {\bibinfo  {journal}
  {Physical Review A}\ }\textbf {\bibinfo {volume} {85}},\ \bibinfo {pages}
  {060301} (\bibinfo {year} {2012})}\BibitemShut {NoStop}%
\bibitem [{\citenamefont {Fujii}\ and\ \citenamefont
  {Tokunaga}(2012)}]{Fujii12a}%
  \BibitemOpen
  \bibfield  {author} {\bibinfo {author} {\bibfnamefont {K.}~\bibnamefont
  {Fujii}}\ and\ \bibinfo {author} {\bibfnamefont {Y.}~\bibnamefont
  {Tokunaga}},\ }\href@noop {} {\bibfield  {journal} {\bibinfo  {journal}
  {Physical Review A}\ }\textbf {\bibinfo {volume} {86}},\ \bibinfo {pages}
  {020303} (\bibinfo {year} {2012})}\BibitemShut {NoStop}%
\bibitem [{\citenamefont {Tomita}\ and\ \citenamefont
  {Svore}(2014)}]{Tomita14a}%
  \BibitemOpen
  \bibfield  {author} {\bibinfo {author} {\bibfnamefont {Y.}~\bibnamefont
  {Tomita}}\ and\ \bibinfo {author} {\bibfnamefont {K.~M.}\ \bibnamefont
  {Svore}},\ }\href@noop {} {\bibfield  {journal} {\bibinfo  {journal}
  {Physical Review A}\ }\textbf {\bibinfo {volume} {90}},\ \bibinfo {pages}
  {062320} (\bibinfo {year} {2014})}\BibitemShut {NoStop}%
\bibitem [{\citenamefont {Darmawan}\ and\ \citenamefont
  {Poulin}(2017)}]{Darmawan17a}%
  \BibitemOpen
  \bibfield  {author} {\bibinfo {author} {\bibfnamefont {A.~S.}\ \bibnamefont
  {Darmawan}}\ and\ \bibinfo {author} {\bibfnamefont {D.}~\bibnamefont
  {Poulin}},\ }\href@noop {} {\bibfield  {journal} {\bibinfo  {journal}
  {Physical review letters}\ }\textbf {\bibinfo {volume} {119}},\ \bibinfo
  {pages} {040502} (\bibinfo {year} {2017})}\BibitemShut {NoStop}%
\bibitem [{\citenamefont {Wang}\ \emph {et~al.}(2011)\citenamefont {Wang},
  \citenamefont {Fowler},\ and\ \citenamefont {Hollenberg}}]{Wang11a}%
  \BibitemOpen
  \bibfield  {author} {\bibinfo {author} {\bibfnamefont {D.~S.}\ \bibnamefont
  {Wang}}, \bibinfo {author} {\bibfnamefont {A.~G.}\ \bibnamefont {Fowler}}, \
  and\ \bibinfo {author} {\bibfnamefont {L.~C.}\ \bibnamefont {Hollenberg}},\
  }\href@noop {} {\bibfield  {journal} {\bibinfo  {journal} {Physical Review
  A}\ }\textbf {\bibinfo {volume} {83}},\ \bibinfo {pages} {020302} (\bibinfo
  {year} {2011})}\BibitemShut {NoStop}%
\bibitem [{\citenamefont {Bombin}\ \emph {et~al.}(2012)\citenamefont {Bombin},
  \citenamefont {Andrist}, \citenamefont {Ohzeki}, \citenamefont {Katzgraber},\
  and\ \citenamefont {Martin-Delgado}}]{Bombin12a}%
  \BibitemOpen
  \bibfield  {author} {\bibinfo {author} {\bibfnamefont {H.}~\bibnamefont
  {Bombin}}, \bibinfo {author} {\bibfnamefont {R.~S.}\ \bibnamefont {Andrist}},
  \bibinfo {author} {\bibfnamefont {M.}~\bibnamefont {Ohzeki}}, \bibinfo
  {author} {\bibfnamefont {H.~G.}\ \bibnamefont {Katzgraber}}, \ and\ \bibinfo
  {author} {\bibfnamefont {M.}~\bibnamefont {Martin-Delgado}},\ }\href@noop {}
  {\bibfield  {journal} {\bibinfo  {journal} {Physical Review X}\ }\textbf
  {\bibinfo {volume} {2}},\ \bibinfo {pages} {021004} (\bibinfo {year}
  {2012})}\BibitemShut {NoStop}%
\bibitem [{\citenamefont {Jouzdani}\ \emph {et~al.}(2013)\citenamefont
  {Jouzdani}, \citenamefont {Novais},\ and\ \citenamefont
  {Mucciolo}}]{Jouzdani13a}%
  \BibitemOpen
  \bibfield  {author} {\bibinfo {author} {\bibfnamefont {P.}~\bibnamefont
  {Jouzdani}}, \bibinfo {author} {\bibfnamefont {E.}~\bibnamefont {Novais}}, \
  and\ \bibinfo {author} {\bibfnamefont {E.}~\bibnamefont {Mucciolo}},\
  }\href@noop {} {\bibfield  {journal} {\bibinfo  {journal} {Physical Review
  A}\ }\textbf {\bibinfo {volume} {88}},\ \bibinfo {pages} {012336} (\bibinfo
  {year} {2013})}\BibitemShut {NoStop}%
\bibitem [{\citenamefont {Novais}\ and\ \citenamefont
  {Mucciolo}(2013)}]{Novais13a}%
  \BibitemOpen
  \bibfield  {author} {\bibinfo {author} {\bibfnamefont {E.}~\bibnamefont
  {Novais}}\ and\ \bibinfo {author} {\bibfnamefont {E.~R.}\ \bibnamefont
  {Mucciolo}},\ }\href@noop {} {\bibfield  {journal} {\bibinfo  {journal}
  {Physical review letters}\ }\textbf {\bibinfo {volume} {110}},\ \bibinfo
  {pages} {010502} (\bibinfo {year} {2013})}\BibitemShut {NoStop}%
\bibitem [{\citenamefont {Jouzdani}\ \emph {et~al.}(2014)\citenamefont
  {Jouzdani}, \citenamefont {Novais}, \citenamefont {Tupitsyn},\ and\
  \citenamefont {Mucciolo}}]{Jouzdani14a}%
  \BibitemOpen
  \bibfield  {author} {\bibinfo {author} {\bibfnamefont {P.}~\bibnamefont
  {Jouzdani}}, \bibinfo {author} {\bibfnamefont {E.}~\bibnamefont {Novais}},
  \bibinfo {author} {\bibfnamefont {I.}~\bibnamefont {Tupitsyn}}, \ and\
  \bibinfo {author} {\bibfnamefont {E.~R.}\ \bibnamefont {Mucciolo}},\
  }\href@noop {} {\bibfield  {journal} {\bibinfo  {journal} {Physical Review
  A}\ }\textbf {\bibinfo {volume} {90}},\ \bibinfo {pages} {042315} (\bibinfo
  {year} {2014})}\BibitemShut {NoStop}%
\bibitem [{\citenamefont {Jouzdani}\ and\ \citenamefont
  {Mucciolo}(2014)}]{Jouzdani14b}%
  \BibitemOpen
  \bibfield  {author} {\bibinfo {author} {\bibfnamefont {P.}~\bibnamefont
  {Jouzdani}}\ and\ \bibinfo {author} {\bibfnamefont {E.~R.}\ \bibnamefont
  {Mucciolo}},\ }\href@noop {} {\bibfield  {journal} {\bibinfo  {journal}
  {Physical Review A}\ }\textbf {\bibinfo {volume} {90}},\ \bibinfo {pages}
  {012315} (\bibinfo {year} {2014})}\BibitemShut {NoStop}%
\bibitem [{\citenamefont {Fowler}(2013)}]{Fowler13a}%
  \BibitemOpen
  \bibfield  {author} {\bibinfo {author} {\bibfnamefont {A.~G.}\ \bibnamefont
  {Fowler}},\ }\href@noop {} {\bibfield  {journal} {\bibinfo  {journal}
  {Physical Review A}\ }\textbf {\bibinfo {volume} {88}},\ \bibinfo {pages}
  {042308} (\bibinfo {year} {2013})}\BibitemShut {NoStop}%
\bibitem [{\citenamefont {Suchara}\ \emph {et~al.}(2015)\citenamefont
  {Suchara}, \citenamefont {Cross},\ and\ \citenamefont
  {Gambetta}}]{Suchara15a}%
  \BibitemOpen
  \bibfield  {author} {\bibinfo {author} {\bibfnamefont {M.}~\bibnamefont
  {Suchara}}, \bibinfo {author} {\bibfnamefont {A.~W.}\ \bibnamefont {Cross}},
  \ and\ \bibinfo {author} {\bibfnamefont {J.~M.}\ \bibnamefont {Gambetta}},\
  }in\ \href@noop {} {\emph {\bibinfo {booktitle} {2015 IEEE International
  Symposium on Information Theory (ISIT)}}}\ (\bibinfo {organization} {IEEE},\
  \bibinfo {year} {2015})\ pp.\ \bibinfo {pages} {1119--1123}\BibitemShut
  {NoStop}%
\bibitem [{\citenamefont {Hanks}\ \emph {et~al.}(2017)\citenamefont {Hanks},
  \citenamefont {Trupke}, \citenamefont {Schmiedmayer}, \citenamefont {Munro},\
  and\ \citenamefont {Nemoto}}]{Hanks17a}%
  \BibitemOpen
  \bibfield  {author} {\bibinfo {author} {\bibfnamefont {M.}~\bibnamefont
  {Hanks}}, \bibinfo {author} {\bibfnamefont {M.}~\bibnamefont {Trupke}},
  \bibinfo {author} {\bibfnamefont {J.}~\bibnamefont {Schmiedmayer}}, \bibinfo
  {author} {\bibfnamefont {W.~J.}\ \bibnamefont {Munro}}, \ and\ \bibinfo
  {author} {\bibfnamefont {K.}~\bibnamefont {Nemoto}},\ }\href {\doibase
  10.1088/1367-2630/aa8085} {\bibfield  {journal} {\bibinfo  {journal} {New
  Journal of Physics}\ }\textbf {\bibinfo {volume} {19}},\ \bibinfo {pages}
  {103002} (\bibinfo {year} {2017})}\BibitemShut {NoStop}%
\bibitem [{\citenamefont {Bhaskar}\ \emph {et~al.}(2019)\citenamefont
  {Bhaskar}, \citenamefont {Riedinger}, \citenamefont {Machielse},
  \citenamefont {Levonian}, \citenamefont {Nguyen}, \citenamefont {Knall},
  \citenamefont {Park}, \citenamefont {Englund}, \citenamefont {Lon\u{c}ar},
  \citenamefont {Sukachev},\ and\ \citenamefont {Lukin}}]{Bhaskar19a}%
  \BibitemOpen
  \bibfield  {author} {\bibinfo {author} {\bibfnamefont {M.~K.}\ \bibnamefont
  {Bhaskar}}, \bibinfo {author} {\bibfnamefont {R.}~\bibnamefont {Riedinger}},
  \bibinfo {author} {\bibfnamefont {B.}~\bibnamefont {Machielse}}, \bibinfo
  {author} {\bibfnamefont {D.~S.}\ \bibnamefont {Levonian}}, \bibinfo {author}
  {\bibfnamefont {C.~T.}\ \bibnamefont {Nguyen}}, \bibinfo {author}
  {\bibfnamefont {E.~N.}\ \bibnamefont {Knall}}, \bibinfo {author}
  {\bibfnamefont {H.}~\bibnamefont {Park}}, \bibinfo {author} {\bibfnamefont
  {D.}~\bibnamefont {Englund}}, \bibinfo {author} {\bibfnamefont
  {M.}~\bibnamefont {Lon\u{c}ar}}, \bibinfo {author} {\bibfnamefont {D.~D.}\
  \bibnamefont {Sukachev}}, \ and\ \bibinfo {author} {\bibfnamefont {M.~D.}\
  \bibnamefont {Lukin}},\ }\href@noop {} {\bibfield  {journal} {\bibinfo
  {journal} {Preprint at \url{https://arxiv.org/abs/1909.01323}}\ } (\bibinfo
  {year} {2019})}\BibitemShut {NoStop}%
\bibitem [{\citenamefont {Wang}\ \emph {et~al.}(2003)\citenamefont {Wang},
  \citenamefont {Harrington},\ and\ \citenamefont {Preskill}}]{Wang03a}%
  \BibitemOpen
  \bibfield  {author} {\bibinfo {author} {\bibfnamefont {C.}~\bibnamefont
  {Wang}}, \bibinfo {author} {\bibfnamefont {J.}~\bibnamefont {Harrington}}, \
  and\ \bibinfo {author} {\bibfnamefont {J.}~\bibnamefont {Preskill}},\
  }\href@noop {} {\bibfield  {journal} {\bibinfo  {journal} {Annals of
  Physics}\ }\textbf {\bibinfo {volume} {303}},\ \bibinfo {pages} {31}
  (\bibinfo {year} {2003})}\BibitemShut {NoStop}%
\bibitem [{\citenamefont {Stephens}(2014)}]{Stephens14a}%
  \BibitemOpen
  \bibfield  {author} {\bibinfo {author} {\bibfnamefont {A.~M.}\ \bibnamefont
  {Stephens}},\ }\href@noop {} {\bibfield  {journal} {\bibinfo  {journal}
  {Physical Review A}\ }\textbf {\bibinfo {volume} {89}},\ \bibinfo {pages}
  {022321} (\bibinfo {year} {2014})}\BibitemShut {NoStop}%
\bibitem [{\citenamefont {Shor}(1996)}]{Shor96a}%
  \BibitemOpen
  \bibfield  {author} {\bibinfo {author} {\bibfnamefont {P.~W.}\ \bibnamefont
  {Shor}},\ }in\ \href@noop {} {\emph {\bibinfo {booktitle} {Proceedings of
  37th Conference on Foundations of Computer Science}}}\ (\bibinfo
  {organization} {IEEE},\ \bibinfo {year} {1996})\ pp.\ \bibinfo {pages}
  {56--65}\BibitemShut {NoStop}%
\bibitem [{Man()}]{Manhattan19a}%
  \BibitemOpen
  \href@noop {} {\enquote {\bibinfo {title} {Manhattan distance, nist
  dictionary of algorithms and data structures},}\ }\bibinfo {howpublished}
  {\url{https://xlinux.nist.gov/dads/HTML/manhattanDistance.html}},\ \bibinfo
  {note} {accessed: 2019-07-30}\BibitemShut {NoStop}%
\bibitem [{Note1()}]{Note1}%
  \BibitemOpen
  \bibinfo {note} {For the larger code distances there is some variability
  unexplained by the sample size. This behaviour is reproducible; we conjecture
  that it arises from the breakdown between error chain and error class
  probabilities, though the question remains open.}\BibitemShut {Stop}%
\bibitem [{\citenamefont {Fowler}\ \emph {et~al.}(2012)\citenamefont {Fowler},
  \citenamefont {Mariantoni}, \citenamefont {Martinis},\ and\ \citenamefont
  {Cleland}}]{Fowler12a}%
  \BibitemOpen
  \bibfield  {author} {\bibinfo {author} {\bibfnamefont {A.~G.}\ \bibnamefont
  {Fowler}}, \bibinfo {author} {\bibfnamefont {M.}~\bibnamefont {Mariantoni}},
  \bibinfo {author} {\bibfnamefont {J.~M.}\ \bibnamefont {Martinis}}, \ and\
  \bibinfo {author} {\bibfnamefont {A.~N.}\ \bibnamefont {Cleland}},\
  }\href@noop {} {\bibfield  {journal} {\bibinfo  {journal} {Physical Review
  A}\ }\textbf {\bibinfo {volume} {86}},\ \bibinfo {pages} {032324} (\bibinfo
  {year} {2012})}\BibitemShut {NoStop}%
\bibitem [{\citenamefont {Wilson}(1927)}]{Wilson27a}%
  \BibitemOpen
  \bibfield  {author} {\bibinfo {author} {\bibfnamefont {E.~B.}\ \bibnamefont
  {Wilson}},\ }\href@noop {} {\bibfield  {journal} {\bibinfo  {journal}
  {Journal of the American Statistical Association}\ }\textbf {\bibinfo
  {volume} {22}},\ \bibinfo {pages} {209} (\bibinfo {year} {1927})}\BibitemShut
  {NoStop}%
\bibitem [{\citenamefont {Kolmogorov}(2009)}]{Kolmogorov09a}%
  \BibitemOpen
  \bibfield  {author} {\bibinfo {author} {\bibfnamefont {V.}~\bibnamefont
  {Kolmogorov}},\ }\href@noop {} {\bibfield  {journal} {\bibinfo  {journal}
  {Mathematical Programming Computation}\ }\textbf {\bibinfo {volume} {1}},\
  \bibinfo {pages} {43} (\bibinfo {year} {2009})}\BibitemShut {NoStop}%
\bibitem [{\citenamefont {Edmonds}(1965)}]{Edmonds65a}%
  \BibitemOpen
  \bibfield  {author} {\bibinfo {author} {\bibfnamefont {J.}~\bibnamefont
  {Edmonds}},\ }\href@noop {} {\bibfield  {journal} {\bibinfo  {journal}
  {Canadian Journal of mathematics}\ }\textbf {\bibinfo {volume} {17}},\
  \bibinfo {pages} {449} (\bibinfo {year} {1965})}\BibitemShut {NoStop}%
\bibitem [{\citenamefont {Gottesman}(1997)}]{Gottesman97a}%
  \BibitemOpen
  \bibfield  {author} {\bibinfo {author} {\bibfnamefont {D.}~\bibnamefont
  {Gottesman}},\ }\href@noop {} {\bibfield  {journal} {\bibinfo  {journal}
  {Preprint at \url{http://arxiv.org/abs/quant-ph/9705052}}\ } (\bibinfo {year}
  {1997})}\BibitemShut {NoStop}%
\bibitem [{\citenamefont {Poulin}\ and\ \citenamefont
  {Chung}(2008)}]{Poulin08a}%
  \BibitemOpen
  \bibfield  {author} {\bibinfo {author} {\bibfnamefont {D.}~\bibnamefont
  {Poulin}}\ and\ \bibinfo {author} {\bibfnamefont {Y.}~\bibnamefont {Chung}},\
  }\href@noop {} {\bibfield  {journal} {\bibinfo  {journal} {Preprint at
  \url{https://arxiv.org/abs/0801.1241}}\ } (\bibinfo {year}
  {2008})}\BibitemShut {NoStop}%
\bibitem [{\citenamefont {Duclos-Cianci}\ and\ \citenamefont
  {Poulin}(2010{\natexlab{a}})}]{Cianci10a}%
  \BibitemOpen
  \bibfield  {author} {\bibinfo {author} {\bibfnamefont {G.}~\bibnamefont
  {Duclos-Cianci}}\ and\ \bibinfo {author} {\bibfnamefont {D.}~\bibnamefont
  {Poulin}},\ }in\ \href@noop {} {\emph {\bibinfo {booktitle} {2010 IEEE
  Information Theory Workshop}}}\ (\bibinfo {organization} {IEEE},\ \bibinfo
  {year} {2010})\ pp.\ \bibinfo {pages} {1--5}\BibitemShut {NoStop}%
\bibitem [{\citenamefont {Duclos-Cianci}\ and\ \citenamefont
  {Poulin}(2010{\natexlab{b}})}]{Cianci10b}%
  \BibitemOpen
  \bibfield  {author} {\bibinfo {author} {\bibfnamefont {G.}~\bibnamefont
  {Duclos-Cianci}}\ and\ \bibinfo {author} {\bibfnamefont {D.}~\bibnamefont
  {Poulin}},\ }\href@noop {} {\bibfield  {journal} {\bibinfo  {journal}
  {Physical review letters}\ }\textbf {\bibinfo {volume} {104}},\ \bibinfo
  {pages} {050504} (\bibinfo {year} {2010}{\natexlab{b}})}\BibitemShut
  {NoStop}%
\bibitem [{\citenamefont {Duclos-Cianci}\ and\ \citenamefont
  {Poulin}(2014)}]{Cianci14a}%
  \BibitemOpen
  \bibfield  {author} {\bibinfo {author} {\bibfnamefont {G.}~\bibnamefont
  {Duclos-Cianci}}\ and\ \bibinfo {author} {\bibfnamefont {D.}~\bibnamefont
  {Poulin}},\ }\href@noop {} {\bibfield  {journal} {\bibinfo  {journal}
  {Quantum Information \& Computation}\ }\textbf {\bibinfo {volume} {14}},\
  \bibinfo {pages} {721} (\bibinfo {year} {2014})}\BibitemShut {NoStop}%
\end{thebibliography}%

\end{document}